\documentclass[aip,reprint,cha]{revtex4-1}   	
\usepackage[pdftex]{graphicx}

\usepackage[usenames,dvipsnames,svgnames,table]{xcolor}
\definecolor{black}{rgb}{0,0,0}
\definecolor{myblue}{rgb}{0,0,0.6}
\definecolor{mygreen}{rgb}{0,0.5,0}
\newcommand{\cmedit}[1]{\textcolor{black}{#1}}
\newcommand{\rgedit}[1]{\textcolor{black}{#1}}

\usepackage{amsmath,amssymb}
\usepackage{comment}
\usepackage{bbold}
\usepackage[caption=false]{subfig}
\usepackage[normalem]{ulem}
\usepackage{tabularx}
\usepackage{multirow}

\newcommand{\ii}{\mathrm{i}}

\newcommand{\arXiv}[1]{{\tt arXiv:#1}}
\newcommand{\ellcrit}{l_{v}}

\newif\ifcolor
\colortrue

\begin{document}

\title{Unstable Spiral Waves and Local
Euclidean Symmetry in a Model of Cardiac Tissue}
\author{Christopher D. Marcotte}
\affiliation{School of Physics, Georgia Institute of Technology, Atlanta, GA 30332, USA}
\author{Roman O. Grigoriev}
\affiliation{School of Physics, Georgia Institute of Technology, Atlanta, GA 30332, USA}
\date{\today}						

\begin{abstract}
This paper investigates the properties of unstable single-spiral wave solutions arising in the Karma model of two-dimensional cardiac tissue. In particular, we discuss how such solutions can be computed numerically \cmedit{on domains of arbitrary shape} and study how their stability, rotational frequency, and spatial drift depend on the size of the domain as well as the position of the spiral core with respect to the boundaries. We also discuss how the breaking of local Euclidean symmetry due to finite size effects as well as the spatial discretization of the model is reflected in the structure and dynamics of spiral waves. This analysis allows identification of a self-sustaining process responsible for maintaining the state of \cmedit{spiral chaos} featuring multiple interacting spirals.
\end{abstract}



\maketitle

\begin{quotation}
Unstable solutions of reaction-diffusion models often inherit the Euclidean symmetries of the underlying evolution equations despite the presence of boundary conditions, or spatial discretization, that break those symmetries. A good example is provided by spiral traveling waves, which arise in models of excitable systems possessing both translational and rotational invariance. In particular, unstable spiral waves are believed to play a major role in initiating and sustaining cardiac arrhythmias such as atrial or ventricular fibrillation. While stable spiral solutions can be readily computed in any geometry, computing unstable spirals can be rather non-trivial, especially when the cores of the spirals drift. We show how unstable spiral waves can be computed on domains of arbitrary shape without making any simplifying assumptions and investigate how the structure and dynamics of these waves depends on the domain size, the position of the spiral core, as well as the relevant parameters of cellular kinetics.
\end{quotation}

\section{Introduction}

Spatially coherent contractions that occur in the heart are essential for pumping blood through the circulatory system. 
The atria, followed by the ventricles, relax, filling with blood, and then contract in a spatially coherent way, pushing the blood out. This coordinated relaxation and contraction, referred to as normal rhythm, can be disrupted by various mechanisms, leading to a range of arrhythmic behaviors, in which the efficiency of the heart as a pump is reduced, sometimes drastically. Fibrillation is the most complex regime characterized by ``turbulent'' dynamics featuring multiple interacting unstable spirals where both spatial and temporal coherence of the contractile dynamics is destroyed. Ventricular fibrillation \cite{Cherry08,Gray1997} is particularly dangerous and, if not treated within minutes, causes death.

Numerous models of cardiac tissue of varying complexity exist. 
The majority of so-called monodomain models fall in the general class of reaction-diffusion systems \cite{Clayton2011, Courtemanche:1996dl, barkley1991model, karma94, Bueno2008, Simitev2006, Noble1962, Cherry2007pv}. Besides cardiac tissue, reaction-diffusion systems are also used to model diverse phenomena such as chemical reactions \cite{Muller:1985th,Keener:1986sp}, bacterial chemotaxis \cite{Siegert91,Vasiev94}, and disease propagation \cite{Murray:1986on}. 
All of these systems display turbulent solutions dominated by multiple interacting spiral waves, though the terminology varies: such solutions are referred to as defect-mediated turbulence \cite{Ouyang1991}, spiral chaos \cite{hagberg1994-from}, spiral breakup \cite{Panfilov1993}, or spiral defect chaos \cite{Morris1993}. 
This lack of consistency reflects the mostly empirical approach to the study of multi-spiral patterns and our lack of fundamental understanding of their dynamics.

Previous attempts to build a dynamical description of fibrillation (and more generally, spiral chaos) were based mostly on the intuition gleaned from the studies of {\it stable} solitary spirals. There is a vast literature on the subject, so we will only mention studies most relevant to the present investigation. In particular, Barkley {\it et al.} \cite{BaKnTu90} showed that even very simple reaction-diffusion models can produce qualitatively different types of spirals. The spiral tip can move either in a circular trajectory or in a periodic or quasi-periodic fashion. By using a reference frame rotating with the spiral, the tip dynamics can be simplified: in the first case the tip becomes stationary while in the second it executes a periodic motion. Respective spirals are referred to as pinned or meandering. \cmedit{Our use of the term ``pinned'' to describe non-drifting spiral waves deviates from the common usage -- in which a spiral wave is attached to macroscopic heterogeneities of the tissue or medium -- only in scale. Since this distinction is superficial, we will use this term to describe spirals pinned to microscopic features as well.}
Barkley \cite{barkley1992} subsequently showed that Newton-Krylov method can be used to efficiently compute pinned spiral waves in a rotating reference frame in which they become stationary (i.e., are described by relative equilibria). He also computed their leading eigenvalues using Arnoldi method \cite{GoOrMa87} and verified that transition to meandering spirals (relative periodic orbits) is described by a Hopf bifurcation. The same approach has later been applied for computing unstable spiral waves in a model of cardiac tissue by Allexandre and Otani \cite{Otani2004}.

\rgedit{Barkley \cite{Barkley94} and Barkley and Kevrekidis \cite{BarKev94} showed that the tip dynamics for both pinned and meandering spirals can be reproduced using a system of five weakly nonlinear ordinary differential equations (ODE) derived by assuming the dynamics are equivariant under a Euclidean symmetry group.} Fiedler {\it et al.} \cite{FiSaScWu96,FiTu98} showed that, more generally, \cmedit{equivariance with respect to noncompact, finite-dimensional Lie groups (such as the Euclidean group $E(n)$) allows description of the dynamics near relative equilibria (such as rigidly rotating spirals) in terms of a skew-product flow, where the motion transverse to the group manifold is decoupled from the motion on the group manifold. Locally, the group manifold represents all symmetry transformations of a particular solution. For instance, in two dimensions, the skew decomposition separates the evolution of the shape of the spiral from the changes in the position or phase of the spiral.} Sandstede {\it et al.} performed a center manifold reduction of the dynamics near relative equilibria \cite{SaScWu97} and near relative periodic orbits \cite{SaScWu99}, formalizing and extending the reduced description of Barkley and Kevrekidis \cite{Barkley94,BarKev94} to physical spaces of arbitrary dimensionality. \cmedit{An in-depth discussion of the role of symmetries in dynamics is available in, e.g., Chossat and Lauterbach \cite{ChossLaut00}.}

Beyn and Th\"ummler~\cite{BeTh04} developed a numerical method for computing the dynamics near rotating relative equilibria on unbounded domains which used the skew-product representation of the dynamics to eliminate or ``freeze'' the dynamics along the group manifold. The freezing approach was later used by Beyn and Lorentz \cite{BeLo08} to numerically compute the entire stability spectra for pinned spiral waves. They also found good agreement between the numerically computed eigenvectors associated with marginal eigenvalues and the Goldstone modes associated with infinitesimal translations and rotations of the spiral wave. The same approach was later used by Hermann and Gottwald \cite{HerGot10} to investigate the dynamics of spiral waves in the large-core limit and by Foulkes and Biktashev \cite{FouBik10} to characterize drift and meandering of spiral waves.

In the case of fibrillation, individual spirals possess rotational frequencies in excess of the normal rhythm pacing~\cite{FentonCherry08ca} and as a result are typically {\it strongly unstable}, often encountering refractory regions of tissue and breaking up within a few rotations \cite{Bar:1993tu,Fenton2002,bernus2003spiral}.
However, at present our understanding of the properties and dynamics of unstable spirals, especially in the context of cardiac dynamics, is limited. This study investigates unstable single-spiral wave solutions in a simple model of cardiac tissue. The key objectives are to understand how their properties are affected by the size of the computational domain, the proximity of the boundaries, and by the spatial discreteness, which is an essential feature of cardiac models reflecting the finite size of cardiac cells -- cardiomyocytes. 

The paper is organized as follows.
We begin in Sect. \ref{sec:Model} with an overview of reaction-diffusion systems and, in particular, the Karma model of cardiac tissue used in the remainder of this study.
The numerical method used to find exact unstable non-chaotic solutions is described in Sect. \ref{sec:NewKry}.  
The results of numerical experiments investigating the properties of unstable single-spiral solutions of the Karma model and the emergence of local Euclidean symmetries are presented in Sect. \ref{sec:Res}.
Sect. \ref{sec:Conc} summarizes these results and discusses their implications in the context of fibrillation \cmedit{(or, more generally, spiral chaos)}.

\section{The model\label{sec:Model}}

Although our focus here is primarily on the dynamics of cardiac tissue, much of the subsequent discussion applies in equal measure to a broader class of excitable systems~\cite{Bikt07} and, even more generally, to reaction-diffusion systems that support wave propagation. Reaction-diffusion equations were originally introduced to describe pattern formation arising in systems involving reacting and diffusing chemicals.
If one forms a vector field from the concentrations of $n$ interacting chemicals, $\mathbf{u}=[u_1,u_2,\cdots,u_n]$, their evolution can be described by the system of partial differential equations (PDE)
\begin{equation}\label{eq:rds}
	\partial_{t}{\mathbf{u}}=D\Delta\mathbf{u}+f(\mathbf{u}),
\end{equation}
\rgedit{where $D$ is a diagonal matrix} of molecular diffusion coefficients, and the two terms on the right-hand-side describe, respectively, diffusion and reactions.

Similar equations, however, have also been used to describe physical and biological systems that do not necessarily involve molecular diffusion. In particular, monodomain models of cardiac tissue dynamics have the same form, although the interpretation of both the variables and the terms on the right-hand-side is different. To distinguish the physical nature of different variables, we will introduce a different notation, $\mathbf{u} = [u,\mathbf{v}]$, where $u$ is an (electric) transmembrane potential and ${\bf v}$ is a set of gating variables. 

Due to the variation between species and even between the different regions of the same heart (e.g., atria, ventricles, Purkinje fibers), there is not a universal model of cardiac dynamics akin to the Navier-Stokes equations for fluid flow. While the equation describing the transmembrane potential, at least in the monodomain formulation, is universal, the models of cellular kinetics vary widely in complexity and even the number of gating variables \cite{FentonCherry08mcc}. Physically more realistic bidomain models \cite{Sepulveda1989} of two- and three-dimensional tissue include an additional nonlocal constraint equation (for the extracellular potential), which makes them significantly more expensive computationally. 
However, their dynamics are qualitatively similar to those of the monodomain models, and when the anisotropy of the intracellular and extracellular media is the same, bidomain models can be reduced to the monodomain formulation \cite{Roth2008}. Hence, our focus here will be entirely on the monodomain model of cardiac tissue dynamics. 

One of the simplest models, which reproduces dynamics qualitatively similar to fibrillation, is due to Karma \cite{karma94}. 
The Karma model was formulated as a minimal model of cardiac dynamics exhibiting alternans -- the instability that is believed to be responsible for initiating and sustaining fibrillation in the heart \cite{Karma1993,karma94}. We use the following nondimensionalized form of the ``reaction'' terms:
\begin{equation}
f(\mathbf{u})=
	\left[\begin{array}{c}
		(u^* - v^{M})\{1 - \tanh(u-3)\}u^{2}/2 - u \\
		\epsilon\left\{\beta \Theta_s(u-1) + \Theta_s(v-1)(v-1) - v\right\}
	\end{array}\right],
\label{eq:modKarma}
\end{equation}
where there is only one gating variable, so that \rgedit{$\mathbf{u} = [u,v]$}.

\begin{figure}[t]
	
  \ifcolor
    \includegraphics[width=\columnwidth]{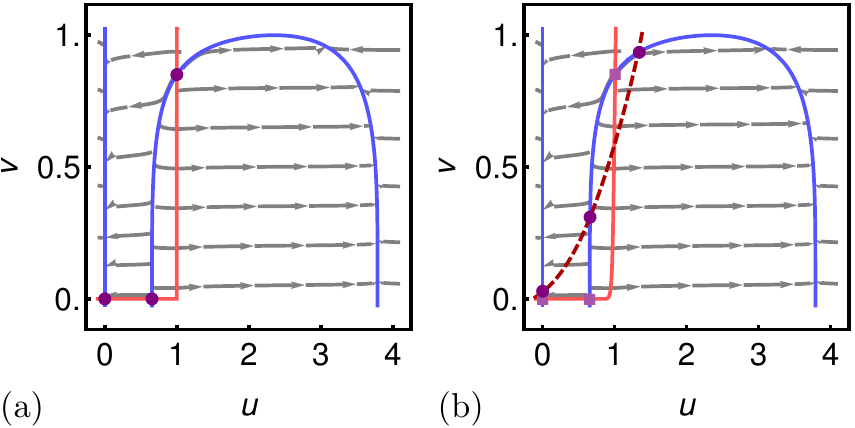}
  \else
    \includegraphics[width=\columnwidth]{figure1_bw.pdf}
  \fi
		
\caption{
The nullclines $\dot{u} = 0$ (blue) and $\dot{v} = 0$ (red) and the vector fields of the original (a) and modified (b) Karma model with $R=1.273$ and $\epsilon=0.01$. \rgedit{In panel (b), the lighter, solid (darker, dashed) curve corresponds to $s=32$ ($s=1.257$). The vector field is shown for $s=32$ and does not change noticeably with $s$. The $u$-nullcline does not vary with $s$.}
\label{fig:1}}
\end{figure}

Several modifications were made to the original model in the interest of creating a well-conditioned dynamical system. All instances of the Heaviside step function were replaced, following Allexandre and Otani \cite{Otani2004}, with a smooth function $\Theta_s(u)=(1+\tanh(su))/2$.
This substitution removes the singularities in the partial derivatives $\partial f_{i}/\partial u_{j}$ while recovering the original step function in the limit $s\rightarrow\infty$. An additional term $\Theta_s(v-1)(v-1)$ was introduced to smooth out an unphysical singularity in the $v$ variable which forms at the spiral core.  In the absence of this term, numerical solutions of the original model become extremely sensitive to the spatial discretization. Finally, the original model assumes no diffusion for the gating variable. In the modified system, we set $D_{22} = \nu D_{11}$ with $0<\nu\ll 1$, to further regularize the dynamics in the core region.
Although diffusion is usually ignored in the equations describing the dynamics of gating variables, all relevant ions and even most secondary messengers such as IP$_3$, cAMP and cGMP can pass through the gap junctions between cells \cite{Bevans98,Garcia04}, so a small nonzero value of $D_{22}$ is justified from the physiological perspective.
These modifications have a very subtle effect on the cellular dynamics. In particular, the $v$-nullcline (Fig. \ref{fig:1}a) is smoothed into a sharply varying, but $C^{\infty}$-continuous function (Fig. \ref{fig:1}b). The position and the stability of the equilibria remain essentially unchanged \rgedit{for $s$ sufficiently large, however, significant deviation in the position of the equilibria is found as $s$ is decreased to $O(1)$.}

We kept the parameter values used in the original work \cite{karma94}, i.e., $u^* = 1.5415$, $M=4$, and $\epsilon=0.01$. All times and lengths were non-dimensionalized using, respectively, the characteristic time scale $\tau_u=2.5$ ms and the characteristic length scale $\xi=0.0262$ cm (the size of a \rgedit{typical} cardiomyocyte).
The original diffusion constant $\gamma=1.1$ cm$^2$/s was non-dimensionalized accordingly, yielding $D_{11} = \gamma\tau_u/\xi^2 = 4.0062$. The ratio of the two diffusion constants was set to $\nu=0.05$. 
\rgedit{Finally, the width $s^{-1}$ of the smoothed step function $\Theta_{s}(u)$ was varied in the range $1<s<32$.}

The dynamics of the modified model with $s\gtrsim 4$ are essentially indistinguishable from those of the original model, so we can rely on the original analysis in choosing the value of the restitution parameter \cmedit{$R=\ln(\beta/(\beta-1))$}. Karma showed \cite{karma94} that the pinned spiral solution is stable at intermediate values of $R$. As $R$ is decreased below about 0.8 (0.6 according to our more precise calculations), it becomes unstable and gives rise to a stable meandering spiral via a Hopf bifurcation. If $R$ is increased above about 1.0, the pinned spiral also undergoes a Hopf instability, which however does not lead to meander. Instead, one finds temporal and spatial modulation of the spiral wavelength (or action potential duration), which is referred to as alternans in the cardiac literature~\cite{Guevara84, Nolasco1968}. 
As $R$ is increased further, conduction block occurs away from the spiral core, leading to a breakup of the spiral and eventual transition to the state of spiral chaos featuring multiple interacting spirals\cmedit{.}
In the remainder of this paper we set $R = 1.273$ (which corresponds to $\beta=1.389$). For this value of $R$ spiral \cmedit{chaos} persists indefinitely for sufficiently large domains \cite{karma94}.

\cmedit{We used} the physiologically and dynamically relevant no-flux boundary conditions, ${\bf n}\cdot\nabla{\bf u} = 0$. These boundary conditions properly describe the vanishing of electric current at the boundaries of cardiac tissue (or the boundaries of its excitable regions). In addition, our investigation \cite{ByMaGr14} of multi-spiral states characterizing \cmedit{spiral chaos} showed that ${\bf n}\cdot\nabla{\bf u}\approx 0$ at the edges of domains (or tiles) supporting individual spirals.

\subsection{Spiral Waves and Symmetry \label{sec:CSandSymm}}

Reaction-diffusion systems \eqref{eq:rds} with arbitrary local kinetics are equivariant with respect to Euclidean transformations ${g} \in G = \mathcal{G} \times E^{+}(1)$, where $\mathcal{G}=E(n)$ for an $n$-dimensional physical space, i.e., $\mathbb{R}^{n}$. The subgroup $E^{+}(1)$ reflects the invariance with respect to time translation, but not time-reversal. In an unbounded two-dimensional domain, the corresponding symmetry group $E(2) \times E^{+}(1)$ includes continuous translations along the two coordinate directions $(x,y)$, continuous rotations in the plane, a discrete inversion symmetry in space, and continuous forward-time translations. As a result, reaction-diffusion systems, and the Karma model in particular, possess solutions that respect some or all of these symmetries on spatially unbounded domains. 

On unbounded domains and in the absence of discretization these symmetries are exact. As a result there are solutions, known as relative equilibria, for which time-evolution is equivalent to a symmetry group transformation. Due to this special property, relative equilibria can be reduced to an equilibrium solution in a moving (e.g., translating or rotating) reference frame. One such example is a rotating, or spiral, wave which satisfies $\partial_{t}\mathbf{u} - \omega\partial_{\theta}\mathbf{u} = 0$, where $\omega$ is the angular speed and $\partial_{\theta}=(x-x_0)\partial_y-(y-y_0)\partial_x$ is the angular derivative about an origin $[x_0,y_0]$. In this particular case, time evolution is equivalent to a rotation about that origin with angular speed $\omega$.

Another, more complex, class of spiral wave solutions is described by relative periodic orbits, i.e., solutions which repeat exactly after some time $T$, $g{\bf u}(T)={\bf u}(0)$, up to a symmetry transformation $g\in\mathcal{G}$. This transformation can include rotations (for meandering spiral waves) or translations (for drifting spiral waves). Similar to relative equilibria, in an appropriate (rotating or translating) reference frame a relative periodic orbit can be reduced to a time-periodic solution.

Spiral wave solutions described by relative equilibria and relative periodic orbits can only be found on unbounded domains, or domains with periodic boundary conditions, since generic boundary conditions break both translational and rotational symmetries (one exception is solutions which respect rotational symmetry relative to the center of a circular domain). Since the motivation for this study is to better understand multi-spiral states, the ability to compute single-spiral solutions on domains of arbitrary shape and size is essential. Furthermore, cardiac tissue is heterogeneous, with the cellular-level models possessing an {\it intrinsic} length scale $\xi$ defined by the size of cardiac cells. This \cmedit{microscopic} heterogeneity also breaks Euclidean symmetry of the PDE (\ref{eq:rds}). 

It is fairly well understood how stable spiral wave solutions change once boundaries are imposed. Spiral wave interaction with boundaries, heterogeneities, etc. is controlled by the size $\ell_c$ of its core, which is determined by the spatial extent of the response functions (adjoint eigenfunctions with unit \rgedit{Floquet multipliers}) \cite{BiHoNi96,BBBBF09,Biktasheva:2010co}. In particular, if the distance between the \cmedit{tip} of the spiral wave and the boundary is larger than $O(\ell_c)$, the structure and dynamics of the wave remain essentially unchanged compared with the corresponding solution on an unbounded domain. However, because of symmetry breaking, a pinned spiral wave would be formally described by a periodic solution, rather than a relative equilibrium, in the presence of boundaries. Similarly, a drifting or meandering wave generally would not be described by a relative periodic orbit in the presence of boundaries. We can expect the same conclusions to also apply to unstable spiral wave solutions.

These complications limit the usefulness of global symmetry reduction as a method for computing spiral wave solutions of cardiac tissue models or characterizing their properties, especially in the unstable regime. Therefore, we developed a different, more general, procedure for computing unstable spiral wave solutions on bounded domains of arbitrary shape by reducing the symmetries locally. It is described in the next section using square domains as a fairly representative example.

\section{Computation of unstable spiral waves \label{sec:NewKry}}

Cardiac tissue is made up of muscle fibers which are, in turn, made up of cardiomyocytes. This structure breaks, perhaps weakly, the rotational symmetry of the system. Hence, even though the choice of coordinate directions in the original PDE \eqref{eq:rds} is arbitrary, the cellular structure imposes a natural choice on the coordinate directions. To be specific, we will assume that the $x$ ($y$) axis is oriented along (transversely to) the fibers.

The modified Karma model \eqref{eq:rds}-\eqref{eq:modKarma} is
solved numerically on square domains of side length $L$ by using a finite-difference discretization of the Laplacian operator on a two-dimensional grid of size $N \times N$. 
Since we chose the size of cardiac cells as \cmedit{the characteristic length scale, $\xi$, by construction} the grid spacing is $\Delta x = \Delta y = 1$ and $L=N$ in non-dimensional units. (In reality the cells are not square, but we will ignore this complication, along with the differences in the diffusion constant along and transversely to the fibers, since these can be accounted for by simply choosing different length scales in the two coordinate directions.) 
The no-flux boundary conditions were enforced using the well-known ``ghost point'' method.

To minimize the impact of discretization on the structure and stability of solutions, the Laplacian operator acting on the state ${\bf u}$ at position $(x,y)$ is approximated using a second-order finite-difference stencil
\begin{equation}\label{eq:FDLap}
	\Delta \mathbf{u}(x,y) \approx \sum_{i,j = -1\dots 1} a_{i,j}
	\mathbf{u}(x + i,y + j),
\end{equation}
\rgedit{with coefficients $a_{\pm1,0} = a_{0,\pm1} = \zeta$, $a_{\pm1,\pm1}=(1-\zeta)/2$, and $a_{0,0} = -2(1+\zeta)$. For $\zeta = 2/3$, this expression is the most isotropic formulation of the nine-point finite-difference stencil on a two-dimensional uniform grid~\cite{lindeberg1990scale}, and is thus expected to most faithfully recover the rotational and translational symmetry of the original PDE.}

The resulting spatially discretized equations are solved numerically using the classical fourth-order Runge-Kutta method, with nondimensional time step $\Delta t = 0.004$. This time step is sufficiently small to accurately resolve the fast time scales in the evolution of the $u$ field, and found to produce solutions accurate to one part in $10^{10}$ in the $2$-norm for integration times of order $10^{4} \Delta t$ (or 100 ms, in dimensional units) with the expected $O(\Delta t^{4})$ convergence, which is sufficient for computing unstable solutions on time scales relevant for the dynamics of this system.

Direct numerical simulation (DNS) can only be used to compute stable (or marginally stable) solutions. Unstable spiral waves described by relative equilibria have been computed with the help of Newton-Krylov method by using rotating reference frames to convert them to equilibria \cite{barkley1992,Otani2004}. However, this approach cannot be extended to finite noncircular domains on which spiral waves are not described by relative equilibria. Recently a new generation of Newton-Krylov methods was developed \cite{Visw07b} that enables efficient computation of many common types of unstable solutions of spatially extended systems, including equilibria and periodic orbits, as well as their relative counterparts in the presence of global continuous symmetries. Specialized versions of these methods applicable to systems with local continuous symmetries are discussed next.

\subsection{Newton-Krylov Solver}

We will illustrate the Newton-Krylov method using relative periodic orbits on an unbounded domain as an example. Consider a drifting spiral wave solution which exactly recurs after a period $T$, but shifted by ${\bf h}=\hat{\bf x}h_x+\hat{\bf y}h_y$ relative to the initial condition (generalization to the case of meandering spiral waves involving rotation is straightforward). Using the evolution operator $\mathcal{U}_T=\exp(T\partial_t)$ and the operator of spatial translations $\mathcal{T}_{\bf h}=\exp({\bf h}\cdot\nabla)$, this can be written as $\mathcal{T}_{-\bf h}\mathcal{U}_T{\bf u}={\bf u}$. This solution can be specified completely by a real-valued vector $\mathbf{w} = [\mathbf{u}, h_x, h_y, T]$, where the state $\mathbf{u} = [u_{1},\dots,u_{N^{2}},v_{1},\dots,v_{N^{2}}]$ is composed of the values of the fields $u$ and $v$ on the computational grid. More generally, a point that lies on a relative periodic orbit corresponds to a root of the vector function ${\bf F}: \mathbb{R}^{2N^{2}} \times \mathbb{R}^{|G|} \rightarrow \mathbb{R}^{2N^{2}}$,
\begin{equation}\label{eq:rpoF}
	\mathbf{F}(\mathbf{w}) =
	g\,\mathcal{U}_T{\bf u} - \mathbf{u},
\end{equation}
which measures the difference between the ``initial'' state $\mathbf{u}$ and the transformed ``final'' state $g\,\mathcal{U}_T{\bf u}$ (here, the more general group transformation $g\in\mathcal{G}$ takes the place of the translation operator $\mathcal{T}_{-\bf h}$). Absolute (non-relative) periodic solutions are merely a special case: they satisfy \eqref{eq:rpoF} with ${g} = \mathbb{1}$, (i.e., ${\bf h} = {\bf 0}$). Similarly, relative equilibria may be treated as special cases of periodic orbits.

Taylor expansion of \eqref{eq:rpoF} shows that the leading-order approximation for the correction $\delta\mathbf{u}$ to the system state is given by the solution of
\begin{equation}\label{eq:dFdw}
    A \, \delta\mathbf{u} = -\mathbf{F}({\bf w}),
\end{equation}
where $A=\partial_{\mathbf{u}}{\bf F}|_{\bf w}$.
However \eqref{eq:dFdw} does not determine the $|G|$ corrections specifying the group transformation and the periodicity, (e.g., the $\delta{\bf h}$ and $\delta T$ of our drifting spiral).
This is a generic feature of systems with continuous symmetries, since there is a continuum of solutions satisfying \eqref{eq:rpoF}. Additional constraints are needed to uniquely specify the solution. Specifically, we will require that the correction to the state $\delta{\bf u}$ be transverse to the group manifold. The constraints can be written using the generators of translation along $x$, $y$, and $t$, which are given by the respective partial derivatives, yielding the following well-posed linear problem:
\begin{equation}\label{eq:linsys}
A'\delta {\bf w}=-{\bf F}',
\end{equation}
where ${\bf F}'=[\mathbf{F}, {\bf 0}]$, and
\begin{equation}
		A'=\begin{pmatrix}
		A		&
		\nabla_{\bf q}\mathbf{u}^{\prime}	\\	
		[\nabla_{\bf q}\mathbf{u}]^{\dagger}	&
		0										\\
	\end{pmatrix},
\end{equation}
where ${\bf u}^{\prime} = g\,\mathcal{U}_T{\bf u}$ denotes the final state, and $\nabla_{\bf q}$ is the \cmedit{row vector of} partial derivatives with respect to the independent coordinates ${\bf q}$ which define the local tangents of the group manifold, e.g., $\nabla_{\bf q}=(\partial_{x},\partial_{y},\partial_{t})$ for the drifting spiral. 

The linear system \eqref{eq:linsys} effectively implements the skew-product representation of the dynamics, where the correction $\delta{\bf w}=[\delta {\bf u},\delta{\bf q}]$ is split into components transverse to, and along, the group manifold. Many authors~\cite{knoll2004} drop the dependence on the final state altogether, instead substituting $\nabla_{\bf q}\mathbf{u}^{\prime} \rightarrow \nabla_{\bf q}\mathbf{u}$. We are unaware of any benefit in doing so and have opted not to, as ${\bf u}'$ is explicitly available: ${\bf u}'={\bf u}+{\bf F}$. 
It should be noted that the transversality condition with respect to rotations is automatically satisfied, since for relative periodic orbits rotations are equivalent to a combination of spatial and temporal translations, obviating the need for an explicit constraint on the ``phase'' of the solution~\cite{barkley1992}. For example, in the special case of relative equilibria we have $\partial_\theta{\bf u}=\omega^{-1}\partial_t{\bf u}$. 

Typical discretizations of \eqref{eq:modKarma} involve $10^4-10^6$ variables, making it very expensive to even compute the elements of the matrix $A'$, let alone solve the system \eqref{eq:linsys} directly.
Instead, the solution can be found approximately using a truncated spectral representation of the matrix, which is constructed using an orthogonal basis produced by Arnoldi iteration \cite{GoOrMa87},
\begin{equation}\label{eq:FDJac}
     \mathbf{v}_{i+1} = A \hat{\mathbf{v}}_{i} \approx
     \frac{\mathbf{F}(\mathbf{w} + \alpha \hat{\mathbf{v}}_{i}) - \mathbf{F}(\mathbf{w})}{\alpha},
\end{equation}
where the action of $A$ is approximated using finite-differencing with some $|\alpha| \ll 1$ and $\hat{\mathbf{v}}_i$ denotes the normalized projection of $\mathbf{v}_i$ onto the orthogonal complement of the linear space spanned by $\{\hat{\mathbf{v}}_1,\cdots, \hat{\mathbf{v}}_{i-1}\}$.

Projecting \eqref{eq:linsys} into the Krylov subspace spanned by $\{\hat{\mathbf{v}}_1,\cdots, \hat{\mathbf{v}}_k\}$ we obtain a $(k+|G|)$-dimensional linear system
\begin{equation}\label{eq:krylov}
    \begin{pmatrix}
        H_k & \mathbf{a}_{\bf q} 	\\
        \mathbf{c}_{\bf q}^{\dagger} & 0	\\
    \end{pmatrix}
    \begin{pmatrix}
    \delta\mathbf{y} \\
    \delta {\bf q} \\
    \end{pmatrix}
    = -\begin{pmatrix}
        \mathbf{b} \\ 0
       \end{pmatrix},
\end{equation}
where $H_k=P_kAP_k^\dagger$ is a real-valued $k\times k$ upper-Hessenberg matrix generated explicitly during the Arnoldi iteration, $\mathbf{c}_{\bf q} = P_k\nabla_{\bf q}\mathbf{u}$, $\mathbf{a}_{\bf q} = P_k\nabla_{\bf q}\mathbf{u}^{\prime}$, $\mathbf{b} = P_k\mathbf{F}$, $\delta\mathbf{y} = P_k\delta\mathbf{u}$, and $P_k=[\hat{\mathbf{v}}_1\cdots \hat{\mathbf{v}}_k]^{\dagger}$ is the projection operator. The vector $\delta {\bf q} = [\delta {\bf h}, \delta T]$ defines the shift and period corrections in our drifting spiral example. Due to strong contraction associated with the Laplacian term in \eqref{eq:rds}, accurate approximation of the original system \eqref{eq:linsys} can usually be obtained with $k$ of order a few tens, very small compared with the full dimensionality of the discretized system, yielding the solution
$\delta\mathbf{w} \approx [P_k^\dagger\delta\mathbf{y}, \delta {\bf q}]$.

Newton-Krylov iterations $\mathbf{w}_{n+1}=\mathbf{w}_n+\delta\mathbf{w}_n$ converge provided a good initial condition ${\bf w}_0$ is available. In order to improve robustness of the procedure for initial conditions that are further away from the solution, a line search was implemented for the cases when the reduction in residual norm was smaller than the expected improvement from linearization, $\|{\bf F}(\mathbf{w}_n+\delta\mathbf{w}_n)\|>\|{\bf F}(\mathbf{w}_n) + A\,\delta\mathbf{w}_{n}\|$. A scaling factor $0<\eta_n \leq 1$ for the Newton step magnitude was computed by minimizing the 2-norm of the residual $\mathbf{F}(\mathbf{w}_n + \eta_n\delta\mathbf{w}_n)$, yielding a more robust sequence of iterations $\mathbf{w}_{n+1}=\mathbf{w}_n+\eta_n\delta\mathbf{w}_n$.
The iteration is considered converged to the solution ${\bf w}$ when the 2-norm of the residual function is minimized below a predetermined value, $\|\mathbf{F}({\bf w})\| < \varepsilon_\mathrm{tol}$ (in this study we used $\varepsilon_\mathrm{tol}= 10^{-10}$).

For pinned single-spiral solutions on arbitrary bounded domains described by absolute periodic orbits the implementation of the Newton-Krylov method described above is sufficient. In fact, it can be simplified by setting ${\bf h}=\delta{\bf h}={\bf 0}$ and dropping the condition of transversality with respect to translations, so that $\delta{\bf q} \to \delta T$. However, Newton-Krylov iterations predictably fail to converge for drifting spirals for which ${\bf h}\ne {\bf 0}$. Since boundary conditions break translational symmetry, the residual \eqref{eq:rpoF} cannot be reduced below $C|{\bf h}|$ for some positive constant $C$.

\subsection{Weighted Newton-Krylov Solver}

\begin{figure}[t]
		
	\ifcolor
	  \includegraphics[width=\columnwidth]{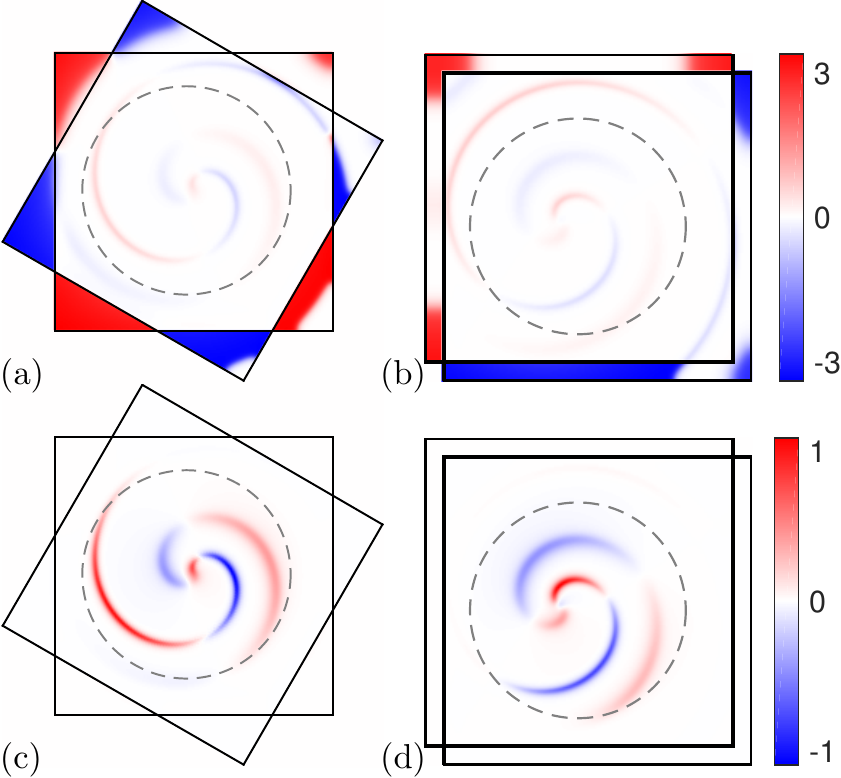}
  \else
    \includegraphics[width=\columnwidth]{figure3_bw.pdf}
  \fi
  
\caption{
The residuals for the spiral waves described by relative periodic orbits. The unweighted residual for a meandering spiral (a) and a drifting spiral (b). The corresponding weighted residuals are shown, respectively, in panels (c) and (d). The $u$ component \rgedit{of the solution} is shown in all the panels. \rgedit{The dashed line corresponds to $r = 0.35L$, which defines the spatial extent of the weighting function.}
\label{fig:3}}
\end{figure}

Boundary conditions do not merely break the translational and rotational symmetries. Finite rotations $\mathcal{R}_\phi=\exp(\phi\partial_\theta)$ on a bounded domain $\Omega$ are not injective: some points are mapped out of the domain and others into (Fig. \ref{fig:3}a). A similar situation is encountered for finite translations $\mathcal{T}_{-\bf h}$ on a bounded domain (Fig. \ref{fig:3}b). Consider, for example, a meandering spiral wave for which $\mathcal{R}_\phi\mathcal{U}_T{\bf u}={\bf u}$ on an unbounded domain. On a bounded domain the residual ${\bf F}=\mathcal{R}_\phi\mathcal{U}_T{\bf u}-{\bf u}$ will not vanish (we set ${\bf u}\equiv 0$ outside of $\Omega$ to make the residual well-defined). If one places the origin of rotation near the \cmedit{tip} of the spiral wave, the residual ${\bf F}$ will decompose into two easily identifiable contributions. Inside $\Omega\cap\mathcal{R}_\phi\Omega$ (the octagonal overlap region in Fig. \ref{fig:3}a), the residual is small and lies near the group manifold,
\begin{equation}\label{eq:grptan}
{\bf F}\approx \delta{\bf q}\cdot\nabla_{\bf q}{\bf u}, 
\end{equation}
where $\delta{\bf q}$ describes the magnitudes of rotations or shifts accounting for the arbitrary choice of the origin and the frequency dependence on the domain size. Outside the overlap region $(\Omega\cup\mathcal{R}_\phi\Omega)\setminus (\Omega\cap\mathcal{R}_\phi\Omega)$ (the eight triangular regions in Fig. \ref{fig:3}a), the residual is large, ${\bf F}=O(1)$.

Figure \ref{fig:3}b describes the effect of boundaries on drifting spiral waves described by relative periodic orbits for which $\mathcal{T}_{-\bf h}\mathcal{U}_T{\bf u}={\bf u}$ on an unbounded domain. On a bounded domain we find a decomposition of the residual ${\bf F}=\mathcal{T}_{-\bf h}\mathcal{U}_T{\bf u}-{\bf u}$ analogous to the case of meandering waves. In the overlap region $\Omega\cap\mathcal{T}_{{\bf h}}\Omega$ (the rectangular region in Fig. \ref{fig:3}b), the residual again can be represented in the form \eqref{eq:grptan}, where the small shifts and rotations account for the dependence of the drift velocity and rotation frequency on the domain size. Outside of the overlap region, again the residual is large, ${\bf F}=O(1)$. Therefore, our objective is to minimize the residual inside the overlap region and suppress it everywhere else.

Generalized relative periodic orbits on bounded domains can be found using a modification of the traditional Newton-Krylov method which introduces an auxiliary weighting (or windowing) of the residual,
\begin{equation}\label{eq:NKphi}
	A' \delta\mathbf{w} = -\rho {\bf F}',
\end{equation}
where $\rho$ is a diagonal matrix with elements \mbox{$0 \leq \rho_{ii} \leq 1$}. The first $2N^{2}$ diagonal elements correspond to the weights associated with the dynamical field variables $u$ and $v$; they correspond to the values of the windowing function
\begin{equation}
W_d({\bf x}) = \frac{1}{2}\left[1 - \tanh\left(\sigma\frac{2|{\bf x}-{\bf x}_c|-dL}{2L}\right)\right],
\end{equation}
where ${\bf x}_c$ denotes the center of the domain and $d$ determines the diameter of the (circular) ``window'' in units of $L$ (we set $d=0.7$ and $\sigma=32$, unless specified otherwise). Specifically, $\rho_{ii}=W_d({\bf x}_i)$. The remaining diagonal elements correspond to the spatial displacements (if applicable) and the period of the spiral wave and are all set to unity. The effect of windowing on the residual is illustrated for the cases of a meandering and a drifting spiral in Fig. \ref{fig:3}c and Fig. \ref{fig:3}d, respectively.
We should point out that the weighting approach is not limited to square domains and rectangular grids and can be easily applied to domains of any shape with any grids, including unstructured ones.

We should also point out the closely related applications of weighting functions in numerical methods for PDEs such as the phase-field boundary method \cite{Bueno2006}, domain decomposition \cite{bjorstad2004domain,xu1992iterative}, and the damping filter method \cite{teramura2014damping}. Alternatively, the weighting may by interpreted as an ad-hoc Jacobi-type preconditioning method for solving linear problems \cite{Trefethen97}. Even though the use of weighting is not a novel numerical approach {\it per se}, to the best of our knowledge, it has never been used either for restoring broken symmetries or for computing unstable traveling wave solutions.

The linear system \eqref{eq:NKphi} can be solved in the same way as \eqref{eq:linsys}. We found that the use of weighting dramatically improves the robustness of the Newton-Krylov method, not only allowing computation of generalized relative periodic orbits, but also substantially improving convergence speed for (absolute) periodic orbits. 
In practice, the convergence properties of the weighted Newton-Krylov solver were found to be fairly insensitive to the shape of the function $W_d({\bf x})$, provided that it vanishes near all the boundaries. In particular, solutions which satisfy \eqref{eq:linsys} can be computed using  \eqref{eq:NKphi} combined with the relaxation process in which $\rho\rightarrow\mathbb{1}$ (or $d\to\infty$). This idea is similar to the damping filter method \cite{teramura2014damping}.

Whether weighting is used or not, the Newton-Krylov solver generates the Floquet multipliers $\Lambda_i$ and Floquet modes ${\bf e}_i$ of the computed solution, i.e, eigenvalues and eigenmodes of the full-space Jacobian $J=\partial_{\bf u}(\mathcal{U}_T{\bf u})$,
\begin{equation}\label{eq:eigen}
    J \mathbf{e}_{i} = \Lambda_{i} \mathbf{e}_{i}
\end{equation}
essentially ``for free.'' Indeed, the spectrum of the Krylov-subspace Jacobian $H_k$ yields a good approximation to the leading eigenvalues and eigenmodes of $A$, while $J={g}^{-1}(A+\mathbb{1})$. Of particular importance are the unstable modes (which correspond to $\Lambda_i>1$) and the Goldstone modes (which correspond to $\Lambda_i=1$) that characterize the symmetries of the system.

\section{Results and Discussion \label{sec:Res}}

\begin{figure}[t]

	\ifcolor
	  \includegraphics[width=\columnwidth]{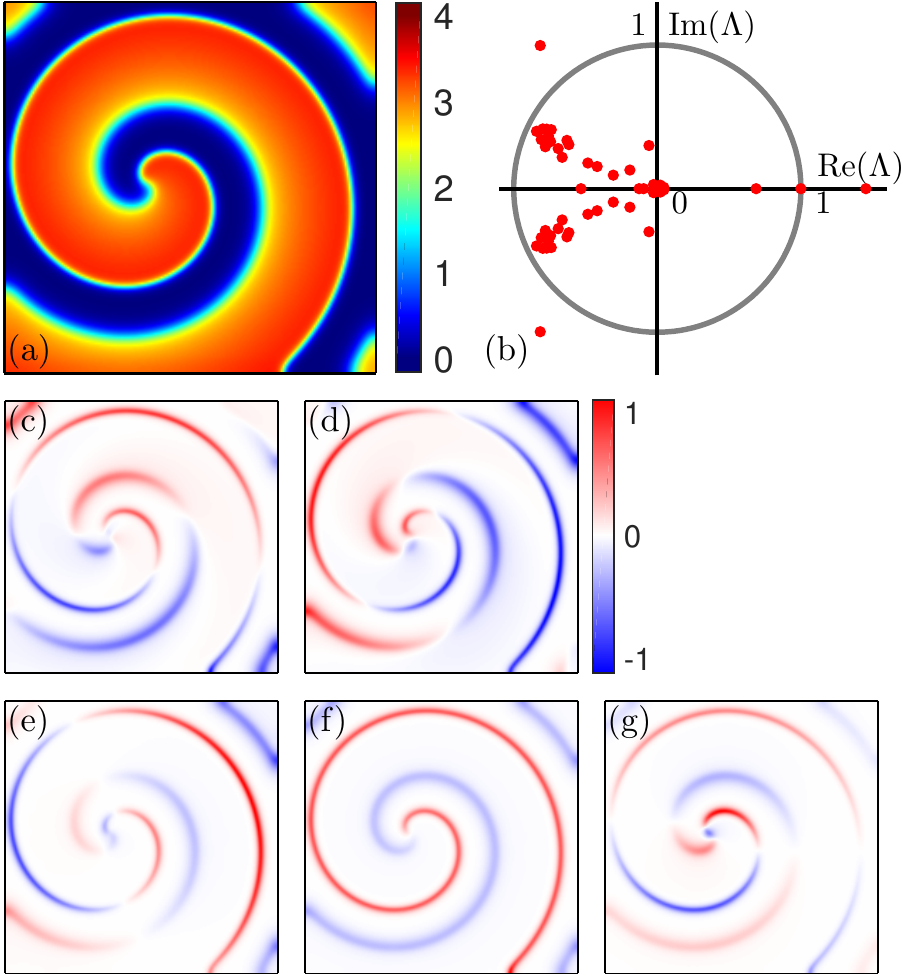}
  \else
    \includegraphics[width=\columnwidth]{figure4_bw.pdf}
  \fi
  
\caption{
A pinned single-spiral solution (a) and the spectrum of its eigenvalues (b) for $s=32$. The real and imaginary parts of the complex-conjugate unstable pair are shown in panels (c) and (d), respectively. 
The color scales shown here are used throughout, unless a different scale is shown. The modes corresponding to the real eigenvalues $\Lambda = 0.69$, $\Lambda=1.00$, and $\Lambda = 1.45$  are shown in panel (e), (f) and (g), respectively. The voltage component $u$ is shown in all the panels. The domain size is $L=192\approx 2.6\lambda$.
\label{fig:4}}
\end{figure}

In this Section we present unstable spiral wave solutions of the modified Karma model computed using the weighted Newton-Krylov method described in Sect. \ref{sec:NewKry} and investigate their properties. In particular, the spiral wave solution shown in Fig. \ref{fig:4}a was found using continuation of a stable single-spiral solution with an approximately centered \cmedit{tip} $[x,y] = [74.43, 96.04]$, achieved by decreasing $\beta$ (which corresponds to increasing the restitution parameter $R$). It has a period $T=50.8273$ and wavelength $\lambda = 74$, or $1.94$ cm in dimensional units. As Fig. \ref{fig:4}b shows, the solution has one Goldstone mode and three unstable modes.

The Goldstone mode presented in Fig. \ref{fig:4}f corresponds to the temporal derivative of the solution $\partial_t{\bf u}$, as expected for a periodic orbit. The real unstable mode (Fig. \ref{fig:4}g) corresponds to an almost rigid shift of the spiral wave in the $y$ direction and can be identified with a frustrated translational Goldstone mode $\partial_y{\bf u}$ (its amplitude varies in space). The stable eigenvalue $\Lambda = 0.6882$ also corresponds to a frustrated Goldstone mode (Fig. \ref{fig:4}e), which can be identified as linear combination of translations in the $x$ and $y$ directions, $\hat{\bf n}\cdot\nabla{\bf u}$, where $\hat{\bf n}$ is the direction of translation. The two complex conjugate unstable modes (Figs. \ref{fig:4}c and \ref{fig:4}d) correspond to the variation in the width of the excitation wave (i.e., alternation of the action potential duration), which is to be expected for $R>1$. Since the Goldstone modes associated with spatial translations are frustrated for $s=32$ (continuous translational symmetry is broken), no additional constraints beyond orthogonality with respect to $\partial_t{\bf u}$ are needed. The residual is minimized for ${\bf h}={\bf 0}$, so this spiral wave is pinned and corresponds to an absolute periodic orbit.

Our results should be contrasted with those obtained by Allexandre and Otani \cite{Otani2004} for the modified version of the 3-variable Fenton-Karma (3V-FK) model \cite{Fenton1998} (they also replaced the Heaviside step functions with smoothed versions $\Theta_s(u)$). The unstable spiral waves of 3V-FK (\cmedit{described by relative equilibria}) were found to possess two near-Goldstone modes corresponding to spatial translations with $\Lambda\approx 1$ and one Goldstone mode corresponding to temporal translation \cmedit{(or spatial rotation)} with $\Lambda=1$. This suggests that translational symmetry is weakly broken due to the presence of boundaries (the calculations were performed on a circular domain of radius equal to just about half the wavelength $\lambda$). In addition, the spectrum included a pair of complex conjugate modes corresponding to meandering instability and a number of unstable modes corresponding to alternans, all laying on the negative real axis (i.e., corresponding to period-doubling, rather than Hopf, bifurcation). We can therefore expect that the alternans instability may lead to substantially different dynamics in the two models.

\begin{figure}[t]

	\ifcolor
	  \includegraphics[width=\columnwidth]{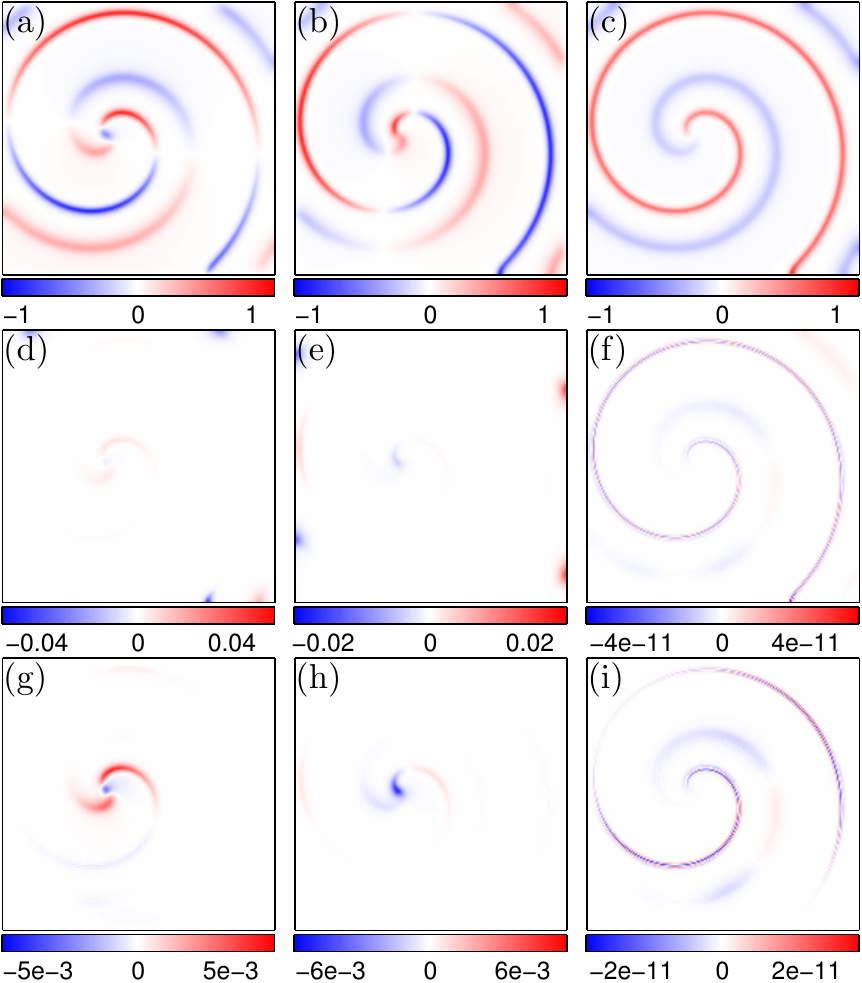}
  \else
    \includegraphics[width=\columnwidth]{figure5_bw.pdf}
  \fi
  
\caption{
Group tangents of the pinned spiral wave solution, $\partial_y u$ (a), $\partial_x u$ (b), and $\partial_t u$ (c). The difference between the group tangents and their images under evolution, $(J-\mathbb{1})\partial_y u$ (d), $(J-\mathbb{1})\partial_x u$ (e), and $(J-\mathbb{1})\partial_t u$ (f).  The windowed difference $W_d(J-\mathbb{1})\partial_y u$ (g), $W_d(J-\mathbb{1})\partial_x u$ (h), and $W_d(J-\mathbb{1})\partial_t u$ (i), where $d=0.9$.
\label{fig:5}}
\end{figure}

The origin of symmetry breaking in the Karma model can be understood by considering the temporal evolution of the group tangents $\partial_x{\bf u}$, $\partial_y{\bf u}$, and $\partial_t{\bf u}$, which become Goldstone modes in the presence of global continuous symmetries. For the Goldstone modes we should have $(J-\mathbb{1}){\bf e}_i=0$ (up to the level of precision defined by $\varepsilon_{\mathrm{tol}}$). We checked this by evolving the group tangents using the linearization of (\ref{eq:rds}) about the periodic solution shown in Fig. \ref{fig:4}a.
As Fig. \ref{fig:5} illustrates, the temporal tangent is a Goldstone mode: $|(J-\mathbb{1})\partial_t{\bf u}|=O(10^{-11})$ \rgedit{while the spatial tangents are not}. Both \mbox{$(J-\mathbb{1})\partial_y{\bf u}$} and $(J-\mathbb{1})\partial_x{\bf u}$ achieve their $O(10^{-2})$ maxima on the boundary of $\Omega$ (Figs. \ref{fig:5}d and \ref{fig:5}e), which confirms the role of boundaries in breaking the global translational symmetry. However, there is another mechanism that also breaks translational symmetry. By applying windowing to suppress the boundary effects, we discover that $\partial_y{\bf u}$ and $\partial_x{\bf u}$ behave like Goldstone modes everywhere except near the core of the spiral wave (Figs. \ref{fig:5}g and \ref{fig:5}h), suggesting that translational symmetry is \cmedit{also} broken locally. We investigate this local mechanism next.

\subsection{Effect of discretization}

As we mentioned previously, spatial discretization can be a local source of symmetry breaking. For the symmetry to get broken the solution should possess a length scale comparable to the scale of spatial discreteness, $\Delta x=\Delta y=1$. The spiral wave solution is characterized by several scales: the wavelength $\lambda= 74$, the length scale $\ell_{u} = \sqrt{D_{11}}\approx 2$ of the fast (voltage) variable, which defines the width of the sharp leading front of the excitation wave, \cmedit{and the length scale $\ellcrit$ over which the dynamics of the slow variable $v$ switches from excitable to refractory. The latter} length scale is controlled by the term $\beta\Theta_s(u-1)$ in \eqref{eq:modKarma} and, to leading order in $\epsilon$, is given by $\ellcrit = 2\ell_{u}/(s\beta)$. Setting $\ellcrit=1$ gives $s = 2\ell_{u}/\beta \approx 3$. Hence, for $\ellcrit\lesssim 1$ ($s\gtrsim 3$) the solution is spatially ill-resolved, and the continuous translational symmetry is broken \cmedit{near the tip}, pinning the spiral. On the other hand, for $\ellcrit\gtrsim 1$ ($s\lesssim 3$) the solution is well-resolved and continuous translational symmetry should be preserved, so the spiral can drift.

\begin{figure}[t]

	\ifcolor
	  \includegraphics[width=\columnwidth]{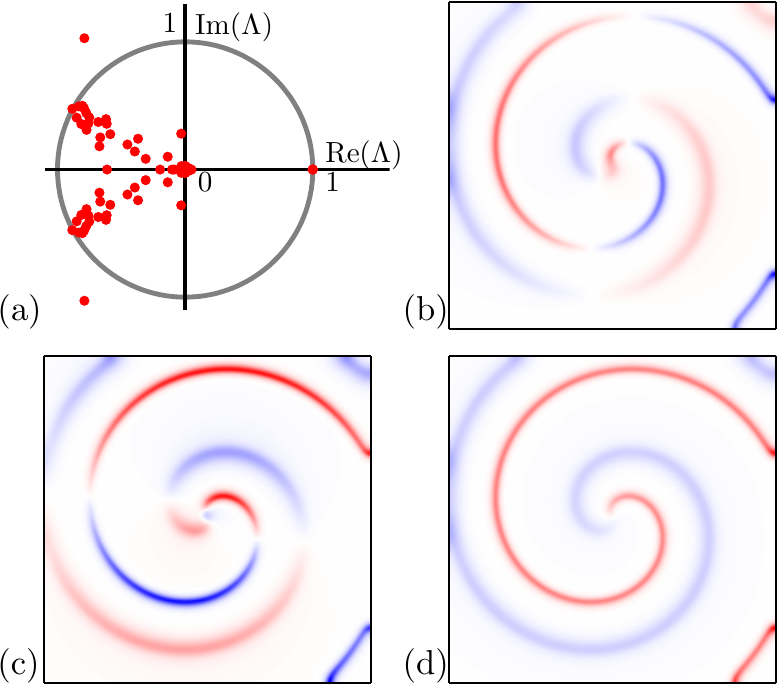}
  \else
    \includegraphics[width=\columnwidth]{figure6_bw.pdf}
  \fi
  
\caption{
The eigenvalues of the drifting single-spiral solution for $s=1.257$ (a). The complex-conjugate pair of unstable modes are similar to those shown in Figs. \ref{fig:4}e and \ref{fig:4}f and thus omitted. The Goldstone modes are shown in panels (b)-(d). The domain size is $L=192$.
\label{fig:6}}
\end{figure}

Indeed, this is exactly what we find for \rgedit{$s = 1.257$}. The spiral wave solution at this low value of $s$ is similar to that shown in Fig. \ref{fig:4}a, but corresponds to a generalized relative periodic orbit with period $T=54.7447$ and wavelength $\lambda=78$. 
The displacement of the wave over one period is small, but non-zero (\rgedit{$|{\bf h}| = O(10^{-9}\lambda)$}), and its direction depends on \rgedit{both the position and the phase} of the spiral wave.
The corresponding spectrum (Fig. \ref{fig:6}a) contains two complex conjugate unstable eigenvalues which correspond to the alternans instability (the corresponding modes are similar to those shown in Figs. \ref{fig:4}e and \ref{fig:4}f) and three eigenvalues with $|\Lambda-1|=O(10^{-6})$. The corresponding Goldstone modes, predictably, coincide with the three group tangents $\partial_t{\bf u}$ (Fig. \ref{fig:6}d), $\partial_y{\bf u}$ (Fig. \ref{fig:6}c), and $\partial_x{\bf u}$ (Fig. \ref{fig:6}b). This indicates that although {\it global} Euclidean symmetry is broken on the finite domain, {\it local} Euclidean symmetry (i.e., symmetry with respect to small translations or rotations) remains essentially exact provided that (i) the domain is sufficiently large and (ii) the length scale of heterogeneities is sufficiently small.

\begin{figure}[t]

  \ifcolor
	  \includegraphics[width=\columnwidth]{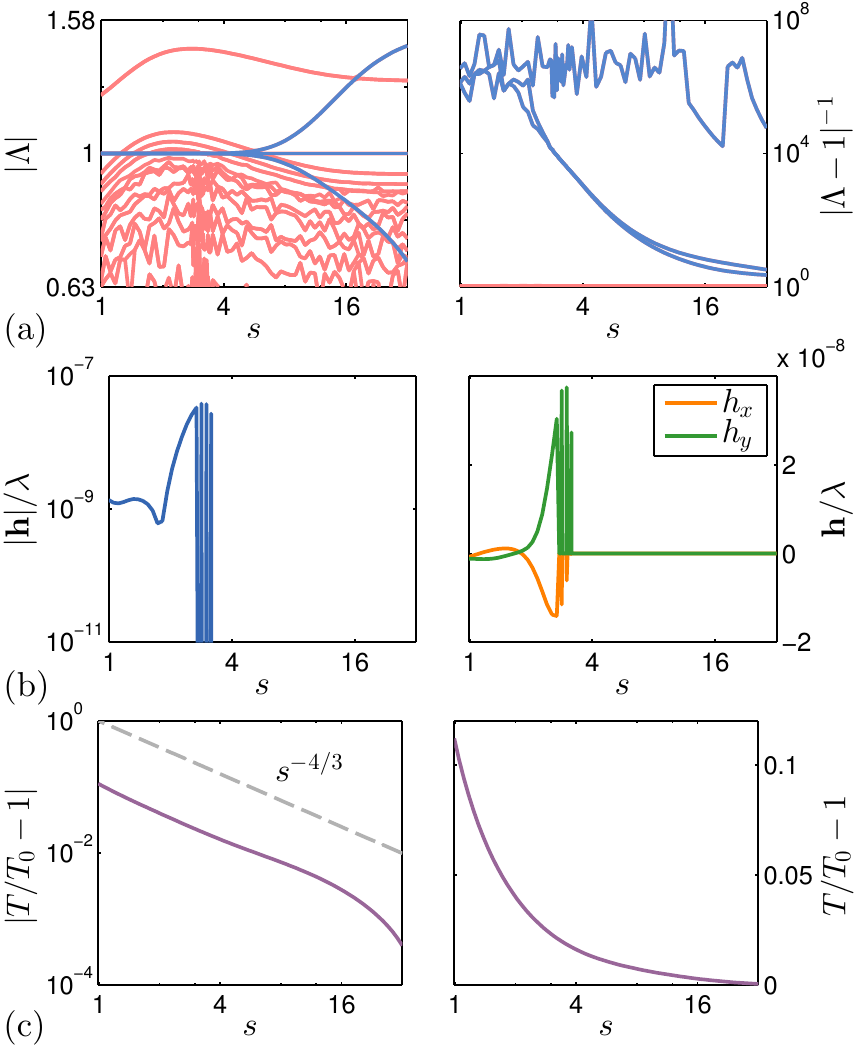}
  \else
    \includegraphics[width=\columnwidth]{figure7_bw.pdf}
  \fi
  
\caption{
Dependence of various properties of the unstable spiral wave solution on the stiffness parameter over the range $1\leq s\leq 32$. (a) Eigenvalues $\Lambda_i$ of unstable and leading stable modes (red) and near-Goldstone modes (blue). (b) The spatial shift ${\bf h}$, normalized by the wavelength $\lambda$. (c) The deviation of the period $T$ from the asymptotic value $T_{0}= 50.8273$.
\label{fig:7}}
\end{figure}

The transition between the small-$s$ regime where local Euclidean symmetry is preserved and the large-$s$ regime where it is broken is continuous. The $s$-dependence of the leading eigenvalues is shown in Fig. \ref{fig:7}a. As the value of $s$ is increased, two of the three unit eigenvalues split off around $s=3$ and separate along the real axis.
For this particular solution, Goldstone mode $\partial_y{\bf u}$ becomes unstable (and frustrated), while the Goldstone mode $\partial_x{\bf u}$ becomes stable (and frustrated) at large $s$.
The exact symmetry of the problem with respect to rotations by $\phi=\pi/2$ means that there is a rotated copy of the solution we found for which the stability of these two modes is interchanged. In either case, however, we find that the $x$ and $y$ directions play a special role: discretization breaks the rotational symmetry despite the use of the Laplacian stencil that aims to preserve it.

It is also worth pointing out that for $1.2 \lesssim s \lesssim 6$, several other modes become unstable. Unlike the modes shown in Figs. \ref{fig:4} and \ref{fig:6} which are isolated (belong to the discrete spectrum), these new modes are not isolated (i.e, belong to the continuous spectrum). Hence, we should expect a continuum of unstable alternans-like modes to appear for single-spiral solutions in unbounded domains for intermediate values of $s$.

The variation of ${\bf h}$, which defines the net spatial displacement of the wave over one period of rotation, with $s$ is shown in Fig. \ref{fig:7}b. Consistent with our dimensional analysis, we find that ${\bf h}$ vanishes for $s\gtrsim 3$, but does not vanish for $s\lesssim 3$. Even though the spirals drift for small $s$, when their \cmedit{tip} is far from domain boundaries, the displacement over one period is too small to be resolved in DNS and can only be computed using extremely precise calculations based on Newton-Krylov method. However, we will see below that the drift can become quite significant for spirals whose \cmedit{tip} is close to a boundary.

The period of the spiral was found to be a monotonically decreasing function of $s$, as Fig. \ref{fig:7}c illustrates. The deviation from the limiting value $T_0$, which corresponds to $s\to\infty$, was found to be well-approximated by a power law, $T-T_0\propto s^{-4/3}$. The overall variation, however, was not large, with the period being just 12\% larger at $s=1$ than at $s=32$.

To further verify our dimensional analysis we also computed the single-spiral solution on progressively finer grids with fixed stiffness parameter $s=32$ and physical domain size $L=192$ and determined that the solution recovers the Goldstone modes associated with translational symmetry on sufficiently fine grids. The unit eigenvalues are recovered for the translational modes with precision $|\Lambda-1|=O(10^{-4})$ when $\Delta x=\Delta y=0.16$, whereas the dimensional analysis predicts a similar critical length scale $\ellcrit=0.09$ for this value of $s$.

\subsection{Boundary effects}

Global translational symmetry implies that there are infinitely many copies of the solution on an unbounded domain that differ in their position but are otherwise equivalent. On bounded domains characterized by local translational symmetry we can also find multiple solutions related to each other by a translation, but they are not, strictly speaking, equivalent. To quantify this relation more precisely we performed a continuation of the domain-centered solutions discussed previously by gradually shifting the \cmedit{tip} towards one of the boundaries on a reasonably large domain. 
The choice of the boundary is arbitrary due to the 4-fold rotational symmetry of the problem.

For the present purposes it is convenient to place the origin of coordinates at the lower left corner of the domain. If the \cmedit{tip} of the spiral (identified as the point at which $\partial_t{\bf u}={\bf 0}$) has coordinates $[x,y]$, then $x$ and $y$ define the distance of the \cmedit{tip} of the spiral, respectively, from the left and bottom boundary. The continuation sequence involves shifting the converged spiral wave solution toward the left or right boundary (decreasing or increasing $x$) to generate an initial condition, which is refined into a new spiral wave solution using the Newton-Krylov solver. This cycle repeats until the Newton-Krylov solver either fails to converge or converges to a previously found solution.

\begin{figure}[t]
  
  \ifcolor
	  \includegraphics[width=\columnwidth]{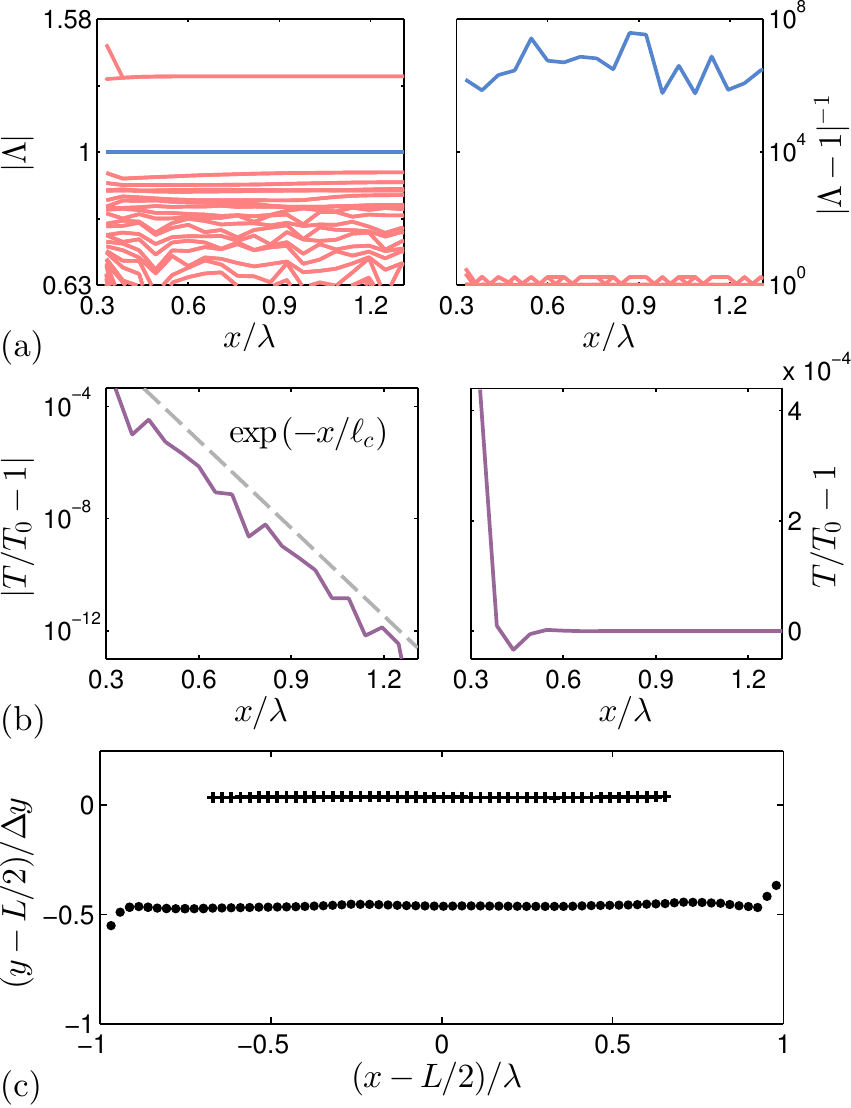}
  \else
    \includegraphics[width=\columnwidth]{figure9_bw.pdf}
  \fi
  
\caption{
Dependence of the properties of pinned spiral waves on the distance to the boundary for $s=32$. (a) The eigenvalues $\Lambda$ 
of unstable and leading stable modes (red) and near-Goldstone modes (blue) of ${\bf u}_0$. (b) The deviation of the period $T$ of ${\bf u}_0$ from the period of the domain-centered solution, $T_{0} = 50.8321$. 
(c) The position of the \cmedit{tip} of the spiral wave for the ${\bf u}_+$ ($+$) and for the ${\bf u}_0$ branch ($\bullet$). \rgedit{Only every other position is shown.}
\label{fig:9}}
\end{figure}

In the large-$s$ regime local {\it continuous} symmetry is broken, so we only find solutions shifted by an integer multiple of the grid spacing $\Delta x$\cmedit{.}
Regardless of the direction (towards the left or right wall) the spiral wave solution \cmedit{with $y\approx L/2$} shown in Fig. \ref{fig:4}a could only be continued for $|x-L/2|\lesssim 0.70\lambda$, \rgedit{i.e., no closer than $0.60\lambda$ to either wall (cf. Fig.~\ref{fig:9}c)}. We will denote this solution branch ${\bf u}_+$. When $x$ was decreased below about $0.60\lambda$, Newton-Krylov solver converged to a nearby spiral wave with $y\approx L/2-\Delta y/2$. This new spiral wave solution can be continued until $x\approx 0.33\lambda$. We will denote this solution branch ${\bf u}_0$. If continued in the opposite direction, ${\bf u}_0$ can be extended symmetrically to $x\approx L-0.33\lambda$. 

\begin{figure}[t]
  
  \ifcolor
	  \includegraphics[width=\columnwidth]{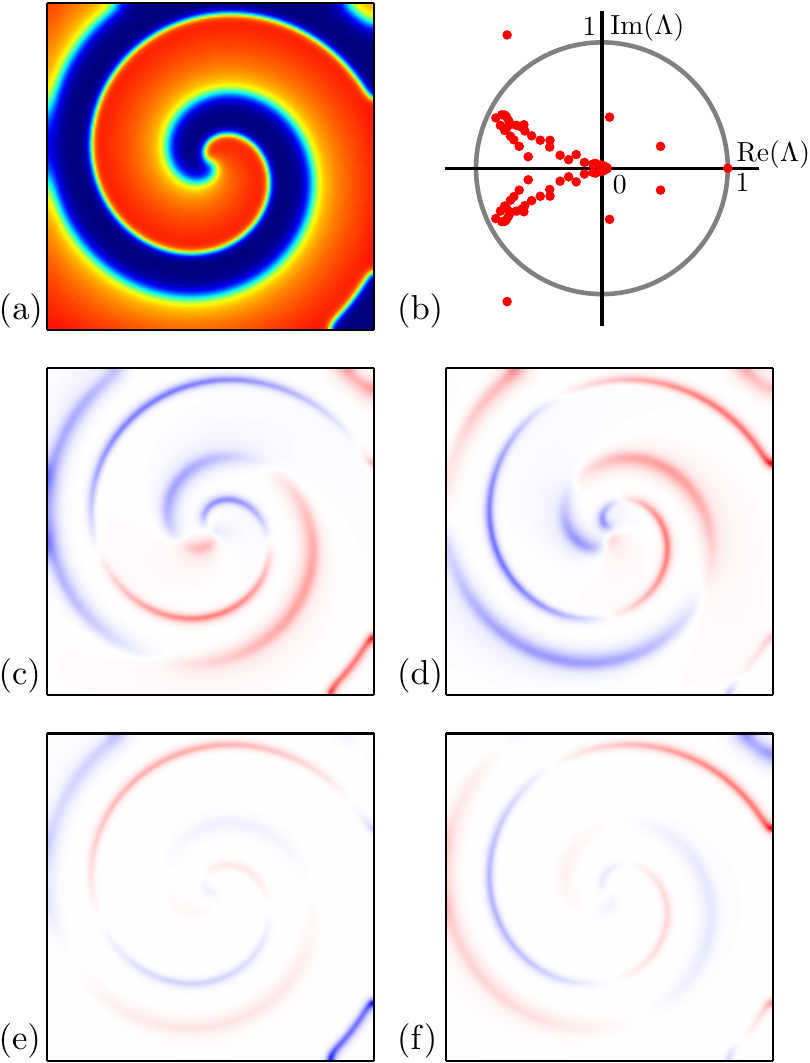}
  \else
    \includegraphics[width=\columnwidth]{figure10_bw.pdf}
  \fi
  
\caption{A sample solution from the branch ${\bf u}_{0}$ (a) and the spectrum of its eigenvalues (b). (c) Real and (d) imaginary part of the unstable complex conjugate pair of modes. (e) Real and (f) imaginary part of the stable complex conjugate pair of modes with eigenvalues $\Lambda_{\pm} = 0.4662 \pm 0.1735\ii$. 
\label{fig:10}}
\end{figure}

An example of a spiral wave corresponding to the branch ${\bf u}_{0}$ with the \cmedit{tip} near $[x,y]=[96.5,95.5]$ along with its spectrum  is shown in Fig.~\ref{fig:10}. Its shape is essentially indistinguishable from that of ${\bf u}_+$ (cf. Fig. \ref{fig:4}a). Its spectrum is also similar to that of ${\bf u}_+$ (cf. Fig. \ref{fig:4}b), except for the eigenvalues with the positive real part: unlike ${\bf u}_+$ which has a pair of  eigenvalues, one unstable and one stable, on the real axis, ${\bf u}_0$ has two stable complex conjugate eigenvalues $\Lambda_{\pm} = 0.4662 \pm 0.1735\ii$. The real and imaginary part of the corresponding modes are shown in Figs. \ref{fig:10}e and \ref{fig:10}f and can be identified as a linear combination of frustrated translational Goldstone modes (their amplitude varies with distance from the \cmedit{tip}).
The real and imaginary part of the complex conjugate pair of unstable modes are shown in Figs. \ref{fig:10}c and \ref{fig:10}d and correspond to the alternans instability.

Counting ${\bf u}_0$, ${\bf u}_+$, and ${\bf u}_-=\mathcal{R}_{\pi/2}{\bf u}_+$, there are at least three distinct pinned spiral wave solutions in the large-$s$ limit. Their \cmedit{tip}s are located, modulo $\Delta x$, approximately at $[0.5,0.5]$ for ${\bf u}_0$, $[0.5,0]$ for ${\bf u}_+$, and $[0,0.5]$ for ${\bf u}_{-}$. Even though these three solutions are distinct, their shapes, temporal periods, and the unstable eigenvalues and eigenmodes corresponding to alternans are virtually indistinguishable, provided they are centered at roughly the same position.
Since they can be shifted in either coordinate direction by an integer multiple of $\Delta x$, there are $O(N^2)$ ``copies'' of each of these three solutions.

\cmedit{Most of these ``copies'' are nearly identical.} Consider, for instance the solution ${\bf u}_0$. Its leading eigenvalues (cf. Fig. \ref{fig:9}a) are effectively independent of the position of the spiral wave and only begin to vary noticeably for $x\lesssim 0.4\lambda$. Similarly, the period of this solution (cf. Fig. \ref{fig:9}b) is essentially independent of the position of the \cmedit{tip} over almost the entire range of $x$. The deviation of the period from the reference value $T_0$ at $x\approx L/2$ decreases exponentially fast with $x$: $|T-T_0|\propto \exp(-x/\ell_c)$, where $\ell_c\approx\lambda/24$ is the characteristic length scale that describes the interaction of this spiral with the boundary.

\begin{figure}[t]
  
  \ifcolor
	  \includegraphics[width=\columnwidth]{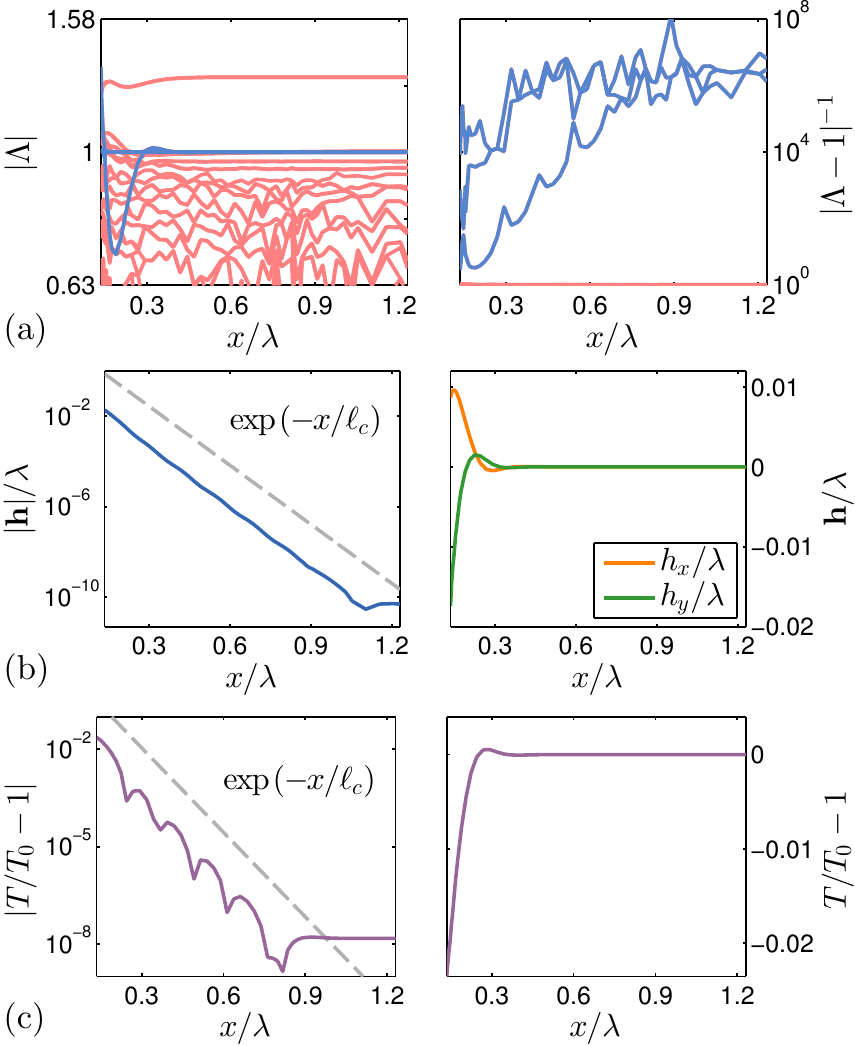}
  \else
    \includegraphics[width=\columnwidth]{figure11_bw.pdf}
  \fi
  
\caption{
Dependence of the properties of drifting spiral waves on the distance to the boundary for $s=1.257$. (a) Eigenvalues $\Lambda_i$ of unstable and leading stable modes (red) and near-Goldstone modes (blue). (b) The spatial shift ${\bf h}$, normalized by the wavelength $\lambda$.
(c) The deviation of the period $T$ from the period of the domain-centered solution, $T_0=54.7446$.
\label{fig:11}}
\end{figure}

In the small-$s$ regime the continuous translational symmetries are recovered, so there is a continuum of drifting spiral wave solutions parametrized by the coordinates ${\bf x} = [x,y]$ of the \cmedit{tip}. Unlike the large-$s$ limit, there is only one type of solution. Its leading eigenvalues $\Lambda$, the spatial shift ${\bf h}$ over one period, and the period $T$ as a function of $x$ (with $y=L/2$) are shown in Fig. \ref{fig:11}. Just like in the large-$s$ limit, we find all the basic properties of the spiral wave solution to be effectively independent of the position of the \cmedit{tip}, provided $x\gtrsim 0.5\lambda$. 
The differences between distinct spiral wave solutions are exponentially small. For instance, Figs. \ref{fig:11}b and \ref{fig:11}c show that $|{\bf h}|\propto\exp(-x/\ell_c)$ and $|T-T_0|\propto\exp(-x/\ell_c)$, where $T_0$ is the period of the centered solution and now $\ell_c\approx\lambda/20$.
The largest differences ($|{\bf h}|\approx 0.05\lambda$ and $|T-T_0|\approx 0.02T_0$) correspond to the distance of the closest approach $x\approx0.14\lambda=O(\ell_c)$.

Our findings can be used to make the concept of local Euclidean symmetry more precise. As long as the \cmedit{tip} of the spiral wave is not too close to the boundary of the domain, that solution can be shifted (discretely for large $s$ or continuously for small $s$) without changing any of its properties up to some level of resolution $\varepsilon_\mathrm{res}$. This level can be made arbitrarily small by increasing the separation between the \cmedit{tip} of the spiral waves and the boundary and is only limited (from below) by the numerical tolerance $\varepsilon_\mathrm{tol}$. Hence, for all practical purposes, distinct spiral wave solutions with their \cmedit{tip}s sufficiently separated from the boundary are completely equivalent, although their spatial shape is \cmedit{affected by the boundaries.}

The symmetry breaking in the proximity of the boundary is reflected in the eigenvalues associated with Goldstone modes in the small-$s$ limit. As Fig. \ref{fig:11}a illustrates, the eigenvalue associated with the $x$-translation deviates from unity for $x\lesssim 0.6\lambda$. 
The vertical boundary does not break the $y$-translation symmetry for spiral waves (and hence does not cause a significant deviation of the corresponding eigenvalue from unity), unless they drift in the $x$ direction.
This symmetry is eventually broken for $x\lesssim 0.3\lambda$ when displacement $h_x$ becomes significant. Comparison with Fig. \ref{fig:11}a shows that for the corresponding eigenvalue $|\Lambda-1|=O(|h_x|/\lambda$).

It is worth noting that \rgedit{$h_x>0$ for $x \lesssim 0.27\lambda$, and $h_x<0$ for $0.27\lambda\lesssim x \lesssim 0.40\lambda$}, so that the interaction is repulsive (the drift is away from the boundary) at small distances and attractive (the drift is towards the boundary) at slightly larger distances. In a related work Langham and Barkley \cite{LanBar13,LanBar14} investigated the interaction with the boundaries for resonantly driven (and hence drifting) {\it stable} spiral waves in the Barkley model \cite{barkley1991model} of excitable media. They also found that the interaction is repulsive at close range and showed that the interaction length scale \rgedit{$\ell_c$ is determined by the spatial extent of the response functions (adjoints of the Goldstone modes) which are localized to the core region~\cite{BiHoBi06,BBBBF09,Biktasheva:2010co}. We have confirmed by direct calculation that both the spatial and temporal response functions in the Karma model (not shown) decay as $e^{-\varrho/\ell_c}$, where $\varrho$ is the distance to the \cmedit{tip}.}

\cmedit{Finally, the period of rotation shows opposite trends for pinned (large $s$) and drifting (small $s$) spirals. As a pinned spiral approaches a boundary it rotates more slowly ($T$ increases), whereas a drifting spiral rotates more quickly ($T$ decreases). A quantitative explanation of these different behaviors relies on the details of the spatial structure of the temporal response function, and is beyond the scope of the current work.}

\subsection{Domain size effects}

We already have some qualitative intuition about the effect the size of the domain has on the structure and properties of spiral wave solutions. As the size of the domain increases, the solution should approach an unbounded spiral and all its properties should become size-independent. In particular, global Euclidean symmetry (whether continuous for small $s$ or discrete for large $s$) should be restored. On the other hand, as the size of the domain is reduced, the structure of the solution and its properties are expected to change significantly, and, on sufficiently small domains the spiral wave solution might not even exist. 

To quantify the differences between single-spiral waves on domains of different size, we computed a sequence of unstable solutions on domains with fixed grid spacing $\Delta x$ and varying physical size $L$\cmedit{.} 
The sequence was initiated from a large spiral wave with an approximately centered core, which was then truncated on all sides by trimming small regions in a symmetric fashion, which generated the initial condition for the next solution. The process continued until the Newton-Krylov solver failed to find a \rgedit{nontrivial solution}.

\begin{figure}[t]
  
  \ifcolor
	  \includegraphics[width=\columnwidth]{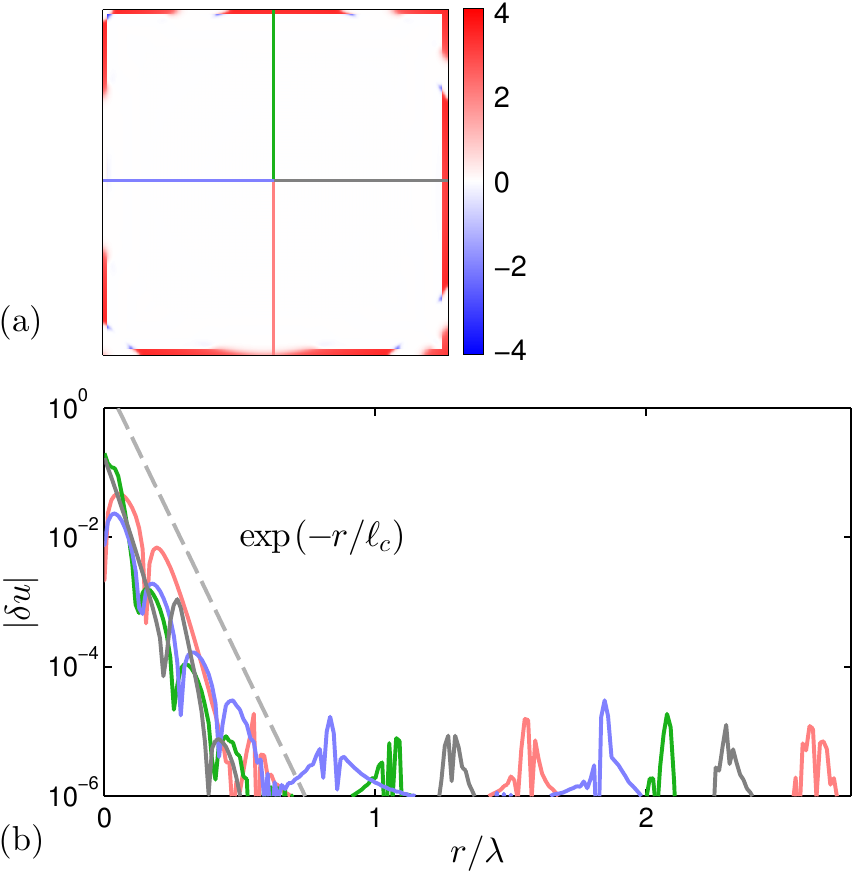}
  \else
    \includegraphics[width=\columnwidth]{figure13_bw.pdf}
  \fi
  
\caption{
The difference $\delta{\bf u}$ between centered spiral wave solutions computed on domains of different size ($L_1=448$ and $L_2=432$) for $s=1.257$. Only the $u$ variable is shown. The results for the $v$ variable are qualitatively similar. (a) The magnitude of the difference in the interior of the smaller domain. (b) The difference along the four rays passing though the \cmedit{tip} of the spirals shown in panel (a) as a function of the distance $r$ from the boundary.
\label{fig:13}}
\end{figure}

Local symmetry suggests that sufficiently far from the boundaries
the solution should not depend on the boundary condition. Hence, solutions computed on domains of different size, e.g., $\Omega_2\subset\Omega_1$, should be virtually indistinguishable away from the boundaries of $\Omega_2$. This is exactly what we find by comparing two solutions ${\bf u}_1$ and ${\bf u}_2$ computed on domains of size $L_1=448$ and $L_2=432$, respectively.
The difference $\delta{\bf u}={\bf u}_1-{\bf u}_2$ inside $\Omega_2$ is found to be concentrated in a narrow boundary layer of width $O(\ell_c)$ (cf. Fig. \ref{fig:13}b), where $\ell_c\approx\lambda/20$ for $s=1.257$, as we determined previously. 
Furthermore, Fig. \ref{fig:13}a shows that the difference becomes exponentially small away from the boundaries, $|\delta u|\propto\exp(-r/\ell_c)$, where $r$ denotes the distance from the boundary of the smaller domain. Inside the boundary layer the solution on the smaller domain adjusts to the no-flux condition and can have a large curvature $\kappa$ comparable to \cmedit{that at the spiral wave} core. Hence, it should not be surprising to see the same length scale describe both the width of the core and the width of the boundary layer.

For $s=32$, the spiral waves are pinned, and we find a minimal domain size, $L_0=17 \approx 0.23\lambda$, approximately $0.45$ cm in dimensional units, below which spiral wave solutions cannot be found. 
This domain size corresponds to the distance between the \cmedit{tip} of the spiral and the boundary equal to $L_0/2=0.115\lambda\approx 2\ell_c\approx \lambda/12$. Since $\ell_c$ determines both the radius of the spiral wave core and the width of the boundary layer, $L=L_0$ corresponds to the collision of the spiral core with the boundary layer. If this criterion applies more generally, we should expect to find spiral waves solutions only on domains of size $L\gtrsim 4\ell_c$. 

\begin{figure}[t]
  
  \ifcolor
	  \includegraphics[width=\columnwidth]{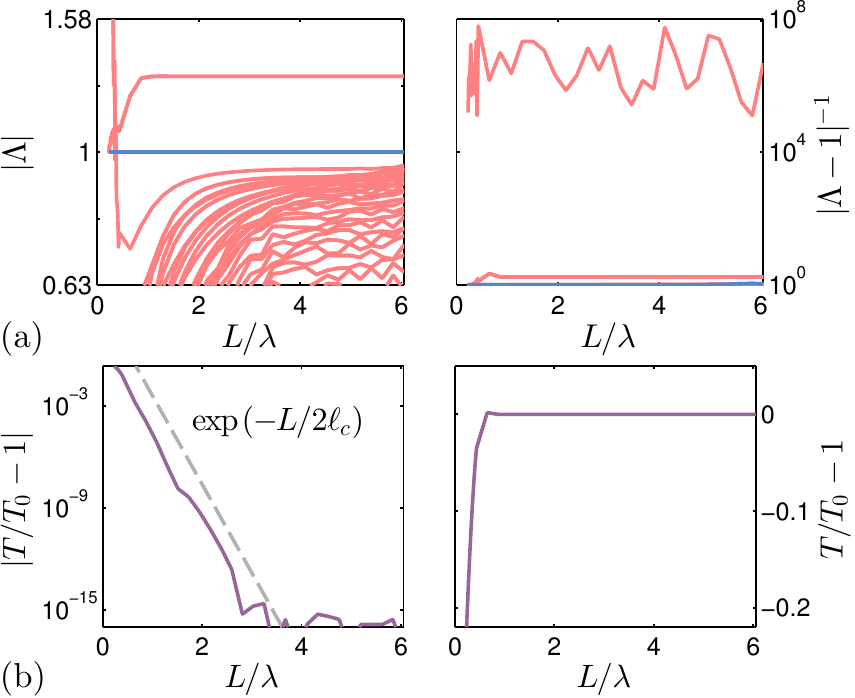}
  \else
    \includegraphics[width=\columnwidth]{figure14_bw.pdf}
  \fi
  
\caption{
Dependence of the properties of the domain-centered pinned spiral wave ${\bf u}_0$ on the size of the domain for $s=32$. (a) The eigenvalues $\Lambda$ of unstable and leading stable modes (red) and 
\cmedit{Goldstone} modes (blue). (b) The deviation of the period $T$ from the period $T_{0} = 50.8321$ at $L=6\lambda$.
\label{fig:14}}
\end{figure}

Pinned spiral waves remain unstable in the entire range of system sizes. The dominant eigenvalues of the solution ${\bf u}_{0}$ are shown in Fig. \ref{fig:14}a.
As $L$ increases, the eigenvalues from the discrete part of the spectrum approach a constant value, but new eigenvalues also appear which correspond to the continuous part of the spectrum. 
For $L_0 \leq L \leq 0.36\lambda$, there are between three and five unstable modes. As $L\to L_0$, the dominant eigenvalues quickly grow in magnitude, approaching $\Lambda_{\pm} = 2.0\pm2.77\ii$ (these are outside of the range of Fig.~\ref{fig:14}a).

\begin{figure}[t]

  \ifcolor
	  \includegraphics[width=\columnwidth]{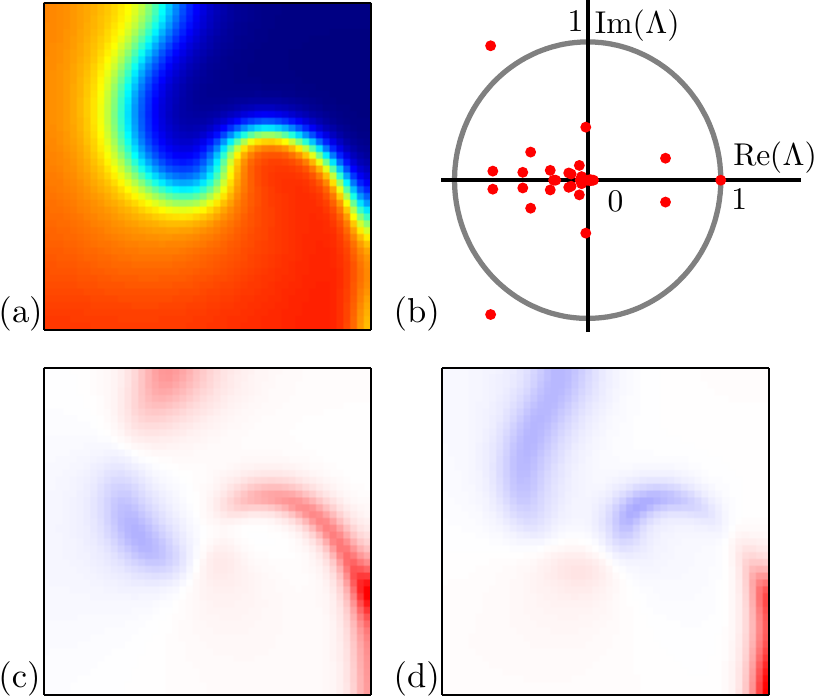}
  \else
    \includegraphics[width=\columnwidth]{figure15_bw.pdf}
  \fi

\caption{
Pinned spiral wave solution on the domain of linear size $L=48\approx 0.65\lambda$ for $s=32$  (a) and its spectrum (b). The real and imaginary parts of the complex conjugate modes corresponding to the unstable eigenvalue pair are shown in panels (c) and (d).
\label{fig:15}}
\end{figure}

\begin{figure}[t]
  
  \ifcolor
	  \includegraphics[width=\columnwidth]{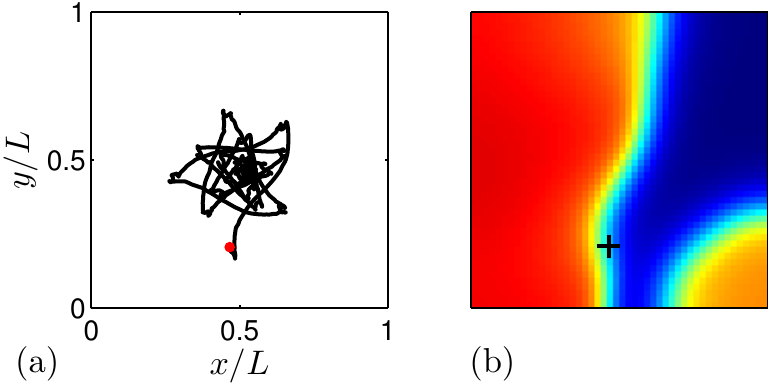}
  \else
    \includegraphics[width=\columnwidth]{figure16_bw.pdf}
  \fi
  
\caption{
Meandering spiral wave solution on the domain of linear size $L=48\approx 0.65\lambda$ for $s=32$. (a) The tip trajectory. (b) The state just before the wave collapse, with \cmedit{tip} marked by the black $+$, which corresponds to the red dot in (a).
\label{fig:16}}
\end{figure}

Interestingly, the character of the instability changes for $L<L_b\approx \lambda$. Consider, for instance, the solution at $L=48\approx 0.65\lambda$ shown in Fig. \ref{fig:15}a. The magnitude of its unstable eigenvalues (cf. Fig. \ref{fig:15}b) is comparable to that in much larger systems, but the corresponding eigenmodes (Fig.~\ref{fig:15}c and \ref{fig:15}d) correspond to the meandering instability, rather than alternans. 
\cmedit{Initial conditions close to this unstable solution produce spiral waves which persist for up to} $10^3$
rotations without breaking up, which is consistent with numerical results of Karma \cite{karma94} for domains \cmedit{smaller than} $2.1$ cm (which corresponds to $L_b=80=1.08\lambda$ in nondimensional units). As. Fig. \ref{fig:16} illustrates, the amplitude of meandering slowly grows, until the tip runs into a boundary and the wave eventually collapses.
The minimal domain size $L_b$ that supports spiral wave breakup via alternans
defines another important dynamical length scale.

The period of the solution approaches the asymptotic value $T_0$ exponentially fast as $L$ increases (cf. Fig.~\ref{fig:14}b). For $L\lesssim 3\lambda$ we find $|T-T_0|\propto\exp(-L/2\ell_c)$,
which is consistent with the scaling we found previously. Indeed, for a domain-centered spiral wave $x=L/2$, or $L=2x$, which means that scaling with $x$ and $L$ follows from the same general scaling law. For $L\gtrsim \lambda$, the period can be considered essentially independent of $L$. 

\begin{figure}[t]
  
  \ifcolor
	  \includegraphics[width=\columnwidth]{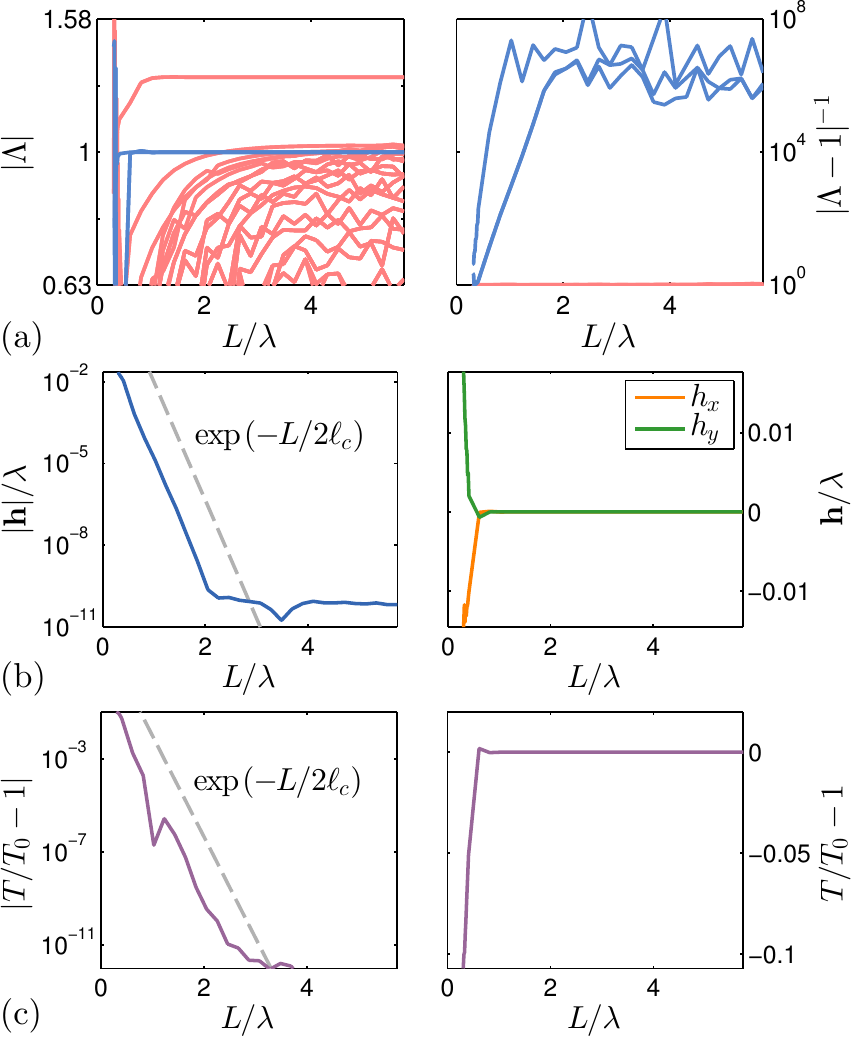}
  \else
    \includegraphics[width=\columnwidth]{figure17_bw.pdf}
  \fi
  
\caption{
Dependence of the properties of a domain-centered drifting spiral wave on the domain size. For $s=1.257$ the minimal domain size is $L_0 = 24 \approx 0.31\lambda$. (a) The eigenvalues $\Lambda$ of unstable and leading stable modes (red) and near-Goldstone modes (blue). (b) The spatial shift ${\bf h}$, normalized by the wavelength $\lambda$. (c) The deviation of the period $T$ from the period $T_{0} = 54.7447$ at $L=5.74\lambda$.
\label{fig:17}}
\end{figure}

Most of the results for small $s$ are qualitatively similar, so we will only focus on what is new or different. This regime is characterized by continuous local symmetries, with drifting spiral waves possessing three Goldstone modes. \rgedit{The magnitude of the spatial displacement of the spiral wave over one period decreases exponentially with $L$, $|{\bf h}|\propto\exp(-L/2\ell_c)$ (cf. Fig.~\ref{fig:17}b). As the domain is truncated, the eigenvalues associated with Goldstone modes} begin to deviate from unity (cf. Fig. \ref{fig:17}a). Eigenvalues corresponding to translational modes deviate first, at $L \approx 1.3\lambda$. This agrees well with the distance $x \approx 0.6\lambda$ to the boundaries at which translational symmetry is broken, as we found in the previous section. Somewhat unexpectedly, for even smaller domains ($L\lesssim 0.6\lambda$), the eigenvalue associated with temporal translation also deviates from unity. \rgedit{The critical domain size matches the distance of the \cmedit{tip} to the boundary ($x \approx 0.3\lambda$) beyond which the spiral solution loses translational symmetry, as noted in the previous section. Hence} deviation of the ``temporal'' eigenvalue from unity can be understood as a result of the non-perturbative nature of {\it finite} spatial shifts on small domains and the subsequent loss of temporal periodicity. \rgedit{For $L\lesssim 0.6\lambda$ the displacement of the spiral is significant enough to affect its temporal dynamics. The solutions in this limit do not describe relative periodic orbits. For instance, both the ``period'' and the spatial shift of ${\bf u}(0)$ and ${\bf u}(T)$ becomes noticeably different.}

The decrease in the temporal period at small separations between the \cmedit{tip} of the spiral and the boundaries observed in Figs. \ref{fig:11}c, \ref{fig:14}b, and \ref{fig:17}c is due to the negative curvature of the wave front near the boundary due to the no-flux boundary condition. It is well-known~\cite{Keener:1986ag} that the speed of propagation of an excitation wave is related to the curvature $\kappa$ of its front by a linear relationship, with convex ($\kappa>0$) wave fronts traveling more slowly than flat ($\kappa=0$) ones, and flat wave fronts traveling more slowly than concave ($\kappa<0$) ones \cite{zykov1979speed,fast1997role}. The magnitude of this effect on the rotation period is controlled by the spatial structure of the temporal response function, which \rgedit{decays} exponentially fast with the distance from the \cmedit{tip} of the spiral, with the decay rate equal to the length scale $\ell_c$.

\rgedit{The exponential dependence of the shift madnitude $|{\bf h}|$ and the period $T$ of the drifting spiral waves found for the modified Karma model are in contradiction with the analytical results obtained by Aranson {\em et. al.}~\cite{aranson1995drift} for a similar model in the $\nu\to 0$ limit, which predict a super-exponential dependence of the drift speed $|{\bf h}|/T$ and rotational frequency $\omega=2\pi/T$ of spiral waves on the domain size $L$. Davydov and Zykov~\cite{davydov1993spiral} predict the frequency of spirals in a generic reaction-diffusion model with $\nu=0$ to vary as the inverse of the domain size (for circular domains of radius comparable to $\lambda$). Hartmann {\em et. al.}~\cite{hartmann1996rotating} claim that their simulations confirm this prediction, but a quick inspection of their numerical results, as well as those of Davydov and Zykov, shows that their data is in excellent agreement with the exponential dependence.}

\section{Summary and discussion\label{sec:Conc}}

We performed the first systematic investigation of unstable spiral wave solutions of a simple spatially discretized model of cardiac tissue with physiologically and dynamically relevant no-flux boundary conditions. We also characterized how the basic properties of these solutions, such as their temporal period, spatial drift, and stability, depend on the size of the domain, the proximity of the spiral core to the nearest domain boundary, and the \cmedit{microscopic} heterogeneity associated with spatial discretization. We found that, although both the boundary conditions and the discretization break the global Euclidean symmetry of the model, unstable spiral wave solutions tend to respect a local -- continuous or discrete -- version of Euclidean symmetry.

Existing \cmedit{tools}, such as Newton-Krylov solvers, developed for computing unstable solutions in the presence of global continuous symmetries (e.g., in unbounded domains or domains with periodic boundary conditions) break down on bounded domains with generic boundary conditions that are not consistent with the global symmetries. However,  for reaction-diffusion systems in general, and monodomain models of cardiac tissue in particular, which lack long-range correlations, the effect of boundaries is localized, which enables computation of solutions satisfying local continuous symmetries using a generalization of Newton-Krylov solvers. The generalization involves weighting, or windowing, of the residual to suppress the symmetry-breaking effect of the boundaries. \cmedit{The generalized solvers permit} the computation of unstable solutions describing, e.g., \cmedit{pinned and} drifting spiral waves in a model of cardiac tissue in the regime characterized by spontaneous breakup of spiral waves leading to fibrillation-like dynamics. 

The inherent spatial heterogeneity associated with the cellular structure of cardiac tissue, and the associated discreteness of the numerical model, was found to have an interesting effect which is usually overlooked in the studies of stable solutions. Cardiac tissue models typically involve discontinuous functions which describe switching of the state of various ionic pumps and channels. Although they simplify the formulation, these discontinuities are unphysical and lead to the emergence of short time scales and corresponding short length scales that can impact the properties of unstable solutions. When these short length scales are small compared to the size of cardiomyocytes, the \cmedit{spatial heterogeneity associated with cellular structure} breaks the continuous translational symmetry, leading to pinning of unstable spiral waves. In the particular cardiac model considered here, at least three different branches of such pinned spiral waves exist, parametrized by the position of their \cmedit{tip} relative to the computational grid and characterized by distinct stability properties. Notably, the \cmedit{``alternans-stable''} meandering spiral waves found in the same parameter regime do not display pinning. However, even there discreteness manifests itself in the jaggedness of the tip trajectory. 

Even though \cmedit{microscopic} spatial heterogeneity breaks the continuous translational symmetry, discrete translation symmetry survives on sufficiently large domains. We find that each unstable solution can be shifted by a discrete number of grid points (or cells) along or transversely to the direction of the fibers without changing, to numerical precision, either its stability properties or its period. Hence, for all practical purposes these solutions can be considered equivalent under discrete translations. This translation symmetry is local in the sense that the properties of different solutions from the same branch are only invariant (to numerical precision) for finite translations such that the \cmedit{tip} of the spiral wave remains outside of the boundary layer of width $O(\ell_c)$, but they begin to vary as the \cmedit{tip} approaches any of the boundaries. For each branch, the translation symmetry is reflected in the stability spectrum in the form of slightly frustrated analogues of translational Goldstone modes characterized by a pair of real or complex conjugate Floquet multipliers with positive real part.
 
Continuous translational symmetry can be restored by replacing the discontinuous switching functions in the ionic model with smooth ones. In the absence of small intrinsic length scales, spiral wave solutions become well-resolved even on a discrete grid. As a result, three discrete branches of pinned spiral wave solutions are merged into a single branch which comprises a continuum of symmetry-related drifting spiral waves parametrized by the position of their \cmedit{tip}. Here too, the continuous translational symmetry is manifested in the stability spectra of the solutions, which possess three Goldstone modes corresponding to the three continuous translation symmetries, one with respect to time and two with respect to spatial \cmedit{position}. Again, the spatial symmetry is local: the properties of spiral wave solutions remain invariant (to numerical precision) with respect to finite translations, provided the \cmedit{tip} of the spiral wave remains outside of the boundary layer of width $O(\ell_c)$.

Global Euclidean symmetry can, of course, be \cmedit{gradually} restored by increasing the size $L$ of the computational domain. However, even on finite domains one finds that the solutions approach their asymptotic shape in the interior of the domain exponentially fast as $L$ increases. The difference between solutions with different $L$ is again found to be confined to the boundary layer of width  $O(\ell_c)$. Similarly, the properties of solutions approach the asymptotic values exponentially fast as $L$ increases, with the same length scale $\ell_c$. This length scale ($\ell_c$) was found to control the effect of the boundaries rather generally. Effectively, $\ell_c$ controls the strength of interaction between a spiral wave and a (locally) straight boundary (and by extension, interaction between two identical counter-rotating spiral waves). With few exceptions (notably, the eigenvalues from the continuous part of the stability spectrum), the deviation from asymptotic values for all properties of spiral wave solutions (e.g., their shapes, temporal periods, spatial displacement over one rotation) were found to scale exponentially, i.e., as $\exp(-r/\ell_c)$, with the distance $r$ to the boundary.

In conclusion, let us comment on the implications of our results for the problem of fibrillation. Unstable spiral wave solutions can only be found for sufficiently large domains, $L\geq L_0$, with the smallest domain size \cmedit{corresponding to the strongly-interacting regime $L_0 \approx 4\ell_c$} in the Karma model. \cmedit{On domains smaller than this size, spiral waves do not persist for a complete rotation.} This length scale determines, in both limits of the stiffness parameter $s$ considered here, the minimal spacing between spiral cores in the state of fibrillation \cite{ByMaGr14}.
As $L$ varies between $L_0$ and $\lambda/2$, the period of the spiral wave varies as much as 20\%, with the smaller spirals rotating \cmedit{faster} than the larger ones. Hence, even if we were to ignore the instabilities on time scales of order a few periods, multi-spiral states with spirals of different size should exhibit dynamics that are very complicated, i.e., at least quasiperiodic.
Quasiperiodicity has been suggested as a possible dynamical mechanism for transition to fibrillation \cite{Garfinkel1997}.

We have established that, \cmedit{for $L<L_b$ meander is the leading instability mechanism, so spirals of size $L_0<L<L_b$ will tend to merge with neighboring spirals of opposite chirality in the same size range. This mechanism reduces the number of spirals and increases the sizes of remaining spirals. On the other hand, 
spirals of size $L>L_b$ are unstable towards alternans and will break up. This mechanism increases the number of spirals by splitting large spirals into smaller ones (ranging in size down to $L_0$).  
The interaction of these two instability mechanisms alone can lead to a dynamic self-sustaining process, which would maintain the state of spiral chaos} featuring multiple interacting spirals ranging in size from $L_0$ to $L_b$. 

\begin{acknowledgments}
The authors would like to thank Vadim Biktashev and Dwight Barkley for helpful discussions and comments. 
This material is based upon work supported by the National Science Foundation under Grant No. CMMI-1028133.  
The Tesla K20 GPUs used for this research were donated by the ``NVIDIA Corporation'' through the academic hardware donation program.
\end{acknowledgments}

\section{References}
\bibliography{../bibtex/cardiac}


\end{document}